\documentclass[letter]{emulateapj}

\usepackage{natbib}
\usepackage{amssymb}
\usepackage{amsmath}
\usepackage{color}
\usepackage{enumitem}
\usepackage{graphicx}

\definecolor{myred}{rgb}{0.8,0.0,0.0}

\newcommand{\rfive}{$R_{500}$}
\newcommand{\mfive}{$M_{500}$}

\newcommand{\yfive}{$Y_{500}$}
\newcommand{\cfive}{$C_{500}$}

\newcommand{\chandra}{{\it Chandra}}
\newcommand{\planck}{{\it Planck}}
\newcommand{\xmm}{{\it XMM}}

\newcommand{\ybig}{$Y_{\textrm{5R500}}$}
\newcommand{\thetas}{$\theta_{\textrm{s}}$}
\newcommand{\thetafive}{$\theta_{500}$}
\newcommand{\boxs}{BoXSZ$^{+}$}
\newcommand{\ymap}{$y$-map}

\begin{document}

\title{A Comparison and Joint Analysis of
  Sunyaev-Zel'dovich Effect Measurements from \planck\ and Bolocam
  for a set of 47 Massive Galaxy Clusters}

\author{
    Jack~Sayers\altaffilmark{1,7},
     Sunil~R.~Golwala\altaffilmark{1},
     Adam~B.~Mantz\altaffilmark{2},
     Julian~Merten\altaffilmark{3},
     Sandor~M.~Molnar\altaffilmark{4},
     Michael~Naka\altaffilmark{1},
     Gregory~Pailet\altaffilmark{1},
     Elena~Pierpaoli\altaffilmark{5},
     Seth~R.~Siegel\altaffilmark{6},
     \& Ben~Wolman\altaffilmark{1},
 }
 \altaffiltext{1}
   {Division of Physics, Math, and Astronomy, California Institute of Technology, 
     Pasadena, CA 91125}
 \altaffiltext{2}
   {Department of Physics, Stanford University, Stanford, CA 94305}
 \altaffiltext{3}
   {Department of Physics, University of Oxford, Oxford OX1 3RH, UK}
 \altaffiltext{4}
   {Institute of Astronomy and Astrophysics, Academia Sinica,
     Taipei 10617, Taiwan}
 \altaffiltext{5}
   {University of Southern California, Los Angeles, CA 90089}
 \altaffiltext{6}
   {Department of Physics, McGill University, Montr{\'e}al, QC H3A 2T8, Canada}
 \altaffiltext{7}
    {jack@caltech.edu}

\begin{abstract}

  We measure the SZ signal toward a set of
  47 clusters with a median mass of $9.5 \times 10^{14}$~M$_{\sun}$ and
  a median redshift of 0.40 using data from \planck\ and the ground-based
  Bolocam receiver.
  When \planck\ \xmm-like masses are used to set the scale radius \thetas,
  we find consistency between the integrated SZ signal,
  \ybig, derived from Bolocam and \planck\
  based on gNFW model fits using A10 shape parameters, with an average
  ratio of $1.069 \pm 0.030$ (allowing for the $\simeq 5$\% Bolocam flux calibration
  uncertainty).
  We also perform a joint fit to the Bolocam and \planck\ data 
  using a modified A10 model with the outer
  logarithmic slope $\beta$ allowed to vary, finding $\beta = 6.13 \pm 0.16 \pm 0.76$
  (measurement error followed by intrinsic scatter).
  In addition, we find that the value of $\beta$ scales with mass and redshift
  according to $\beta \propto M^{0.077 \pm 0.026} \times (1+z)^{-0.06 \pm 0.09}$.
  This mass scaling is in good agreement with recent simulations.
  We do not observe the strong trend of $\beta$ with redshift seen in simulations,
  though we conclude that this is most likely due to our sample selection.
  Finally, we use Bolocam measurements of \yfive\
  to test the accuracy of the \planck\ completeness estimate.
  We find consistency, with the actual number of \planck\ detections
  falling approximately $1 \sigma$ below the expectation from Bolocam.
  We translate this small difference into a constraint on the
  the effective mass bias for the \planck\ cluster cosmology results, with
  $(1-b) = 0.93 \pm 0.06$.

\end{abstract}

\section{Introduction}

  The Sunyaev-Zel'dovich (SZ) effect has emerged as a valuable observational tool
  for studying galaxy clusters, particularly with the dramatic improvements
  in instrumentation that have occurred over the past decade. 
  For example, the South Pole Telescope
  \citep[SPT,][]{Bleem2015}, the Atacama Cosmology Telescope 
  \citep[ACT,][]{Hasselfield2013},
  and \planck\ \citep{Planck2015_XXVII}
  have delivered catalogs with a combined total of more than
  1000 SZ-detected clusters.
  Beyond these large surveys, detailed studies of the 
  gaseous intra-cluster medium (ICM) have been enabled
  by an additional set of pointed SZ facilities with broad spectral 
  coverage and/or excellent angular resolution such as the
  Multiplexed SQUID/TES Array at Ninety GHz \citep[MUSTANG,][]{Mason2010}
  the New IRAM KID Arrays \citep[NIKA,][]{Adam2016}, and
  the Multiwavelength Submillimeter Inductance Camera \citep[MUSIC,][]{Sayers2016}. 
  
  As the range of SZ instrumentation has become more diverse, the
  benefits of joint analyses using multiple datasets have increased.
  For example, a wide range of studies have used data from two
  or more SZ receivers in order to measure the spectral
  shape of the SZ signal
  \citep[e.g.,][]{Kitayama2004, Zemcov2010, Mauskopf2012}, 
  mainly for the purpose of constraining
  the ICM velocity via the kinetic SZ signal, but also to
  characterize relativistic corrections to the classical SZ 
  spectrum
  \citep[e.g.,][]{Sunyaev1980, Nozawa1998, Chluba2012}.
  Furthermore, recent analyses have begun to exploit the 
  different angular sensitivities of the SZ facilities in
  order to obtain a more complete spatial picture of the 
  cluster
  \citep[e.g.,][]{Romero2015, Young2015, Rodriguez-Gonzalvez2016}.

  In order for these joint analyses to be useful, the various
  SZ instruments must provide measurements of the SZ signal
  that are consistent.
  Historically, this was often not the case, likely due to large
  systematic errors in the measurements 
  \citep[e.g., see the detailed discussion in][]{Birkinshaw1999}.
  However, the situation has improved considerably with
  advances in modern SZ instrumentation,
  and good agreement has been seen in most recent comparisons
  \citep[e.g.,][]{Reese2012, Mauskopf2012, Rodriguez-Gonzalvez2016, Sayers2016}.
  Modest inconsistencies do still appear, although they are
  often the result of assuming different spatial templates
  when performing the SZ analyses for separate instruments
  \citep[e.g.,][]{Benson2004, Planck2013_II, Perrott2015}.  
  In sum, the systematics that plagued early SZ measurements
  appear to be largely absent from modern data.
  This fact, combined with the high degree of complementarity between
  different SZ facilities, has opened a promising future
  for detailed cluster studies using multiple SZ datasets.

  In this work, we use SZ measurements from \planck\ and the ground-based
  receiver Bolocam to study a set of 47 massive clusters. The manuscript
  is organized as follows.
  In Section~\ref{sec:SZ_effect}, the parametric model used to describe the
  data is introduced, and in Section~\ref{sec:data} the SZ data from
  \planck\ and Bolocam are detailed.
  Section~\ref{sec:SZ_measurements} compares the SZ signals measured by \planck\
  and Bolocam, and Section~\ref{sec:planck_bolocam_a10} presents the
  results from joint fits to the two datasets.
  In Section~\ref{sec:implications}, we use Bolocam SZ measurements to perform a 
  test of the \planck\ cluster survey completeness, and 
  a summary of the manuscript is given in Section~\ref{sec:summary}.

\section{The SZ Effect}
  \label{sec:SZ_effect}

  The thermal SZ effect \citep{Sunyaev1972}
  describes the Compton scattering of CMB photons
  with hot electrons in the ICM according to
  \begin{eqnarray*}
    \Delta T(\nu) & = & f(\nu,T_{\mathrm{e}}) y \\
    y      & = & \int \frac{k_{\textrm{B}} \sigma_{\textrm{T}}}{m_{\mathrm{e}} c^2} p_{\mathrm{e}} dl 
  \end{eqnarray*}
  where $\Delta T(\nu)$ is the observed surface brightness 
  fluctuation in units of CMB temperature at
  the frequency $\nu$, $T_{\textrm{e}}$ is the ICM electron temperature,
  $f(\nu, T_{\textrm{e}})$ describes the spectral dependence of the SZ signal
  including relativistic corrections~\citep[e.g.,][]{Rephaeli1995, 
    Itoh1998, Nozawa1998, Itoh2004, Chluba2012}, $y$ is the
  SZ Compton parameter, $k_{\textrm{B}}$ is Boltzmann's constant,
  $\sigma_{\textrm{T}}$ is the Thompson cross section, 
  $m_{\textrm{e}}$ is the electron mass, $c$ is the speed of light,
  $p_{\textrm{e}}$ is the ICM electron pressure, and $dl$ is along the line of sight.
  In the absence of relativistic corrections, which are generally small
  and/or constrained using a spectroscopic X-ray measurement
  of the value of $T_{\textrm{e}}$, the SZ brightness gives a 
  direct measure of the integrated ICM pressure.
  Therefore, SZ measurements are often used to constrain parametric
  models of the pressure, such as the generalized Navarro, Frenk,
  and White (gNFW, \citealt{Navarro1997}) model described in
  the following section.

  \subsection{The gNFW Model}
  \label{sec:gnfw}

  \citet{Nagai2007} proposed the use of a gNFW model to describe
  cluster pressure profiles according to
  \begin{displaymath}
    P(R) = \frac{P_0}{(R/R_{\textrm{s}})^{\gamma}
      (1 + (R/R_{\textrm{s}})^{\alpha})^{(\beta - \gamma)/\alpha}}
  \end{displaymath}
  where $P(R)$ is the pressure as a function of radius,
  $P_0$ is the normalization factor, $R_{\textrm{s}}$ is the scale
  radius, and
  $\alpha$, $\beta$, and $\gamma$ control the logarithmic slope
  of the profile at $R \sim R_{\textrm{s}}$, $R \gg R_{\textrm{s}}$,
  and $R \ll R_{\textrm{s}}$.
  Often, the radial coordinates are rescaled
  to angular coordinates denoted by $\theta$ and \thetas,
  and $R_{\textrm{s}}$ is often recast in terms of a
  concentration parameter, with 
  \begin{displaymath}
    C_{500} = R_{500} / R_{\textrm{s}} = \theta_{500} / \theta_{\textrm{s}},
  \end{displaymath}
  and \rfive\ denoting the radius where the average enclosed density is 500
  times the critical density of the universe.
  Therefore, for a given value of \cfive, the values of 
  $R_{\textrm{s}}$ and \thetas\ are directly related to the cluster
  mass, \mfive.
  Furthermore, the normalization is often given in terms the SZ observable
  integrated within a specific radius, for example
  \begin{displaymath}
    Y_{5\textrm{R}500} = \int^{5 \times \theta_{500}}_{0} y \times 2 \pi \theta d\theta. 
  \end{displaymath}
  \citet{Nagai2007} noted that, when $P_0$ is scaled according to a factor 
  that depends on the cluster's mass and redshift and $R_{\textrm{s}}$
  is recast in terms of $C_{500}$, that a single
  set of values for $\alpha$, $\beta$, and $\gamma$ provide
  an approximately universal description of any cluster's pressure
  profile.
  Subsequently, several groups have published different
  values for these logarithmic slopes based on different samples, 
  data, and analysis techniques 
  (e.g., \citealt{Arnaud2010, Plagge2010, Planck2013_V, Sayers2013_pressure, McDonald2014}
  and \citealt{Mantz2016}), and the values
  given by \citet{Arnaud2010} are the most widely used.
  The corresponding gNFW shape with $C_{500} = 1.18$, 
  $\alpha = 1.05$, $\beta = 5.49$, and $\gamma = 0.31$
  is often referred to as the A10 model.

\section{Data}
  \label{sec:data}

  \subsection{Cluster Sample}

    This study focuses on a set of 47 clusters with publicly available
    data from Bolocam\footnote{
      \url{http://irsa.ipac.caltech.edu/data/Planck/release_2/ancillary-data/bolocam/} }
    and \chandra.
    Data for 45 of these clusters were published in \citet{Czakon2015}, who
    named that sample the Bolocam X-ray SZ (BoXSZ) sample.
    Throughout this work, the slightly expanded set of 47 clusters
    is referred to as the \boxs\ sample (see Table~\ref{tab:sample}).
    Based on the \planck\ MMF3 detection algorithm, 32 \boxs\ clusters
    were detected by \planck, with 25 detected at a high
    enough significance to be included in the \planck\
    cluster cosmology analysis \citep{Planck2015_XXIV, Planck2015_XXVII}.

    \begin{deluxetable*}{ccccccccc}
  \tablewidth{0pt}
  \tablecaption{Cluster Sample}
  \tablehead{\colhead{} & \colhead{} & \colhead{RA} & \colhead{dec} & 
    \colhead{\mfive} & \colhead{\thetafive\ ({\it CXO})} & \colhead{\thetafive\ (\xmm)} & 
    \colhead{\planck} & \colhead{Bolocam} \\[-1.2ex]
    \colhead{Cluster} & \colhead{$z$} & \colhead{} & \colhead{} & 
    \colhead{} & \colhead{} & \colhead{} & \colhead{} & \colhead{} \\[-1.2ex]
    \colhead{} & \colhead{} & \colhead{hr} & \colhead{deg} & 
    \colhead{$10^{14}$ M$_{\odot}$} & \colhead{arcmin} & 
    \colhead{arcmin} & \colhead{SNR} & \colhead{SNR}}
  \startdata
    Abell 2204          & 0.15 & 16:32:47 & $+$05:34:32 &    $10.3 \pm 1.5$ & $9.2 \pm 0.4$
     &    $8.6 \pm 0.2$ &    16.3 &    22.3
    \\
    Abell 1689          & 0.18 & 13:11:29 & $-$01:20:27 &    $10.5 \pm 1.5$ & $7.9 \pm 0.4$
     &    $7.6 \pm 0.2$ &    16.7 & \phn6.2
    \\
    Abell 0383          & 0.19 & 02:48:03 & $-$03:31:46 & \phn$4.7 \pm 0.8$ & $5.8 \pm 0.3$
     &    --- &     --- & \phn9.6
    \\
    Abell 0209          & 0.21 & 01:31:53 & $-$13:36:48 &    $12.6 \pm 1.9$ & $7.4 \pm 0.4$
     &    $6.6 \pm 0.2$ &    17.1 &    13.9
    \\
    Abell 0963          & 0.21 & 10:17:03 & $+$39:02:52 & \phn$6.8 \pm 1.0$ & $6.0 \pm 0.3$
     &    $5.8 \pm 0.2$ & \phn8.8 & \phn8.3
    \\
    Abell 1423          & 0.21 & 11:57:17 & $+$33:36:39 & \phn$8.7 \pm 2.0$ & $6.5 \pm 0.5$
     &    $5.7 \pm 0.2$ & \phn9.7 & \phn5.8
    \\
    Abell 2261          & 0.22 & 17:22:26 & $+$32:07:58 &    $14.4 \pm 2.6$ & $7.4 \pm 0.4$
     &    $6.1 \pm 0.2$ &    13.5 &    10.2
    \\
    Abell 0267          & 0.23 & 01:52:42 & $+$01:00:29 & \phn$6.6 \pm 1.1$ & $5.5 \pm 0.3$
     &    $5.0 \pm 0.3$ & \phn5.4 & \phn9.6
    \\
    Abell 2219          & 0.23 & 16:40:20 & $+$46:42:29 &    $18.9 \pm 2.5$ & $7.9 \pm 0.4$
     &    $6.7 \pm 0.1$ &    26.3 &    11.1
    \\
    RX J2129.6$+$0005   & 0.24 & 21:29:39 & $+$00:05:17 & \phn$7.7 \pm 1.2$ & $5.6 \pm 0.3$
     &    $4.6 \pm 0.3$ & \phn4.8 & \phn8.0
    \\
    Abell 1835          & 0.25 & 14:01:01 & $+$02:52:40 &    $12.3 \pm 1.4$ & $6.3 \pm 0.3$
     &    $5.8 \pm 0.2$ &    14.4 &    15.7
    \\
    Abell 0697          & 0.28 & 08:42:57 & $+$36:21:56 &    $17.1 \pm 2.9$ & $6.5 \pm 0.4$
     &    $5.6 \pm 0.1$ &    18.9 &    22.6
    \\
    Abell 0611          & 0.29 & 08:00:56 & $+$36:03:25 & \phn$7.4 \pm 1.1$ & $4.7 \pm 0.2$
     &    $4.3 \pm 0.2$ & \phn6.8 &    10.8
    \\
    Abell 2744          & 0.31 & 00:14:15 & $-$30:23:31 &    $17.6 \pm 3.0$ & $6.0 \pm 0.5$
     &    $4.9 \pm 0.1$ &    14.1 &    15.9
    \\
    MACS J2140.2$-$2339 & 0.31 & 21:40:15 & $-$23:39:40 & \phn$4.7 \pm 0.6$ & $3.9 \pm 0.1$
     &    --- &     --- & \phn6.5
    \\
    Abell S1063         & 0.35 & 22:48:44 & $-$44:31:45 &    $22.2 \pm 3.4$ & $5.9 \pm 0.3$
     &    $4.8 \pm 0.1$ &    20.7 &    13.6
    \\
    MACS J1931.8$-$2635 & 0.35 & 19:31:49 & $-$26:34:33 & \phn$9.9 \pm 1.6$ & $4.5 \pm 0.2$
     &    $3.9 \pm 0.2$ & \phn6.1 &    10.1
    \\
    MACS J1115.8$+$0129 & 0.36 & 11:15:51 & $+$01:29:54 & \phn$8.6 \pm 1.2$ & $4.2 \pm 0.2$
     &    $3.8 \pm 0.2$ & \phn7.1 &    10.9
    \\
    MACS J1532.8$+$3021 & 0.36 & 15:32:53 & $+$30:20:58 & \phn$9.5 \pm 1.7$ & $4.3 \pm 0.3$
     &    --- &     --- & \phn8.0
    \\
    Abell 0370          & 0.38 & 02:39:53 & $-$01:34:38 &    $11.7 \pm 2.1$ & $4.5 \pm 0.3$
     &    $3.9 \pm 0.1$ & \phn7.6 &    12.8
    \\
    MACS J1720.2$+$3536 & 0.39 & 17:20:16 & $+$35:36:22 & \phn$6.3 \pm 1.1$ & $3.6 \pm 0.2$
     &    $3.6 \pm 0.2$ & \phn6.5 &    10.6
    \\
    MACS J0429.6$-$0253 & 0.40 & 04:29:36 & $-$02:53:05 & \phn$5.8 \pm 0.8$ & $3.4 \pm 0.2$
     &    --- &     --- & \phn8.9
    \\
    MACS J2211.7$-$0349 & 0.40 & 22:11:45 & $-$03:49:42 &    $18.1 \pm 2.5$ & $5.0 \pm 0.2$
     &    $4.1 \pm 0.1$ &    11.8 &    14.7
    \\
    ZwCl 0024.0$+$1652  & 0.40 & 00:26:35 & $+$17:09:40 & \phn$4.4 \pm 1.6$ & $3.1 \pm 0.3$
     &    --- &     --- & \phn3.3
    \\
    MACS J0416.1$-$2403 & 0.42 & 04:16:08 & $-$24:04:13 & \phn$9.1 \pm 2.0$ & $3.8 \pm 0.5$
     &    $1.5 \pm 0.2$ & \phn4.7 & \phn8.5
    \\
    MACS J0451.9$+$0006 & 0.43 & 04:51:54 & $+$00:06:18 & \phn$6.3 \pm 1.1$ & $3.3 \pm 0.2$
     &    --- &     --- & \phn8.1
    \\
    MACS J0417.5$-$1154 & 0.44 & 04:17:34 & $-$11:54:27 &    $22.1 \pm 2.7$ & $4.9 \pm 0.2$
     &    $4.0 \pm 0.1$ &    13.3 &    22.7
    \\
    MACS J1206.2$-$0847 & 0.44 & 12:06:12 & $-$08:48:05 &    $19.2 \pm 3.0$ & $4.7 \pm 0.2$
     &    $3.9 \pm 0.1$ &    13.3 &    21.7
    \\
    MACS J0329.6$-$0211 & 0.45 & 03:29:41 & $-$02:11:46 & \phn$7.9 \pm 1.3$ & $3.4 \pm 0.2$
     &    --- &     --- &    12.1
    \\
    MACS J1347.5$-$1144 & 0.45 & 13:47:30 & $-$11:45:08 &    $21.7 \pm 3.0$ & $4.8 \pm 0.2$
     &    $3.8 \pm 0.1$ &    11.2 &    36.6
    \\
    MACS J1311.0$-$0311 & 0.49 & 13:11:01 & $-$03:10:39 & \phn$3.9 \pm 0.5$ & $2.6 \pm 0.1$
     &    --- &     --- & \phn9.6
    \\
    MACS J2214.9$-$1400 & 0.50 & 22:14:57 & $-$14:00:11 &    $13.2 \pm 2.3$ & $3.8 \pm 0.2$
     &    $3.3 \pm 0.1$ & \phn8.3 &    12.6
    \\
    MACS J0257.1$-$2325 & 0.51 & 02:57:09 & $-$23:26:03 & \phn$8.5 \pm 1.3$ & $3.2 \pm 0.2$
     &    $2.9 \pm 0.1$ & \phn5.4 &    10.1
    \\
    MACS J0911.2$+$1746 & 0.51 & 09:11:10 & $+$17:46:31 & \phn$9.0 \pm 1.2$ & $3.3 \pm 0.2$
     &    $2.9 \pm 0.1$ & \phn5.1 & \phn4.8
    \\
    MACS J0454.1$-$0300 & 0.54 & 04:54:11 & $-$03:00:50 &    $11.5 \pm 1.5$ & $3.4 \pm 0.2$
     &    $3.1 \pm 0.1$ & \phn7.1 &    24.3
    \\
    MACS J1149.6$+$2223 & 0.54 & 11:49:35 & $+$22:24:04 &    $18.7 \pm 3.0$ & $4.0 \pm 0.2$
     &    $3.2 \pm 0.1$ &    11.3 &    17.4
    \\
    MACS J1423.8$+$2404 & 0.54 & 14:23:47 & $+$24:04:43 & \phn$6.6 \pm 0.9$ & $2.9 \pm 0.1$
     &    --- &     --- & \phn9.4
    \\
    MACS J0018.5$+$1626 & 0.55 & 00:18:33 & $+$16:26:13 &    $16.5 \pm 2.5$ & $3.8 \pm 0.2$
     &    $3.1 \pm 0.1$ & \phn8.6 &    15.7
    \\
    MACS J0717.5$+$3745 & 0.55 & 07:17:32 & $+$37:45:20 &    $24.9 \pm 2.7$ & $4.4 \pm 0.2$
     &    $3.4 \pm 0.1$ &    12.8 &    21.3
    \\
    MACS J0025.4$-$1222 & 0.58 & 00:25:29 & $-$12:22:44 & \phn$7.6 \pm 0.9$ & $2.8 \pm 0.1$
     &    --- &     --- &    12.3
    \\
    MS 2053             & 0.58 & 20:56:21 & $-$04:37:48 & \phn$3.0 \pm 0.5$ & $2.1 \pm 0.2$
     &    --- &     --- & \phn5.1
    \\
    MACS J0647.8$+$7015 & 0.59 & 06:47:49 & $+$70:14:55 &    $10.9 \pm 1.6$ & $3.2 \pm 0.2$
     &    $2.7 \pm 0.1$ & \phn5.8 &    14.4
    \\
    MACS J2129.4$-$0741 & 0.59 & 21:29:25 & $-$07:41:31 &    $10.6 \pm 1.4$ & $3.1 \pm 0.2$
     &    --- &     --- &    15.2
    \\
    MACS J0744.9$+$3927 & 0.70 & 07:44:52 & $+$39:27:27 &    $12.5 \pm 1.6$ & $2.9 \pm 0.1$
     &    --- &     --- &    13.3
    \\
    CL J1052.7$-$1357   & 0.83 & 01:52:41 & $-$13:58:06 & \phn$7.8 \pm 3.0$ & $2.1 \pm 0.6$
     &    --- &     --- &    10.2
    \\
    MS 1054             & 0.83 & 10:56:58 & $-$03:37:33 & \phn$9.0 \pm 1.3$ & $2.3 \pm 0.2$
     &    --- &     --- &    17.4
    \\
    CL J1226.9$+$3332   & 0.89 & 12:26:57 & $+$33:32:48 & \phn$7.8 \pm 1.1$ & $2.1 \pm 0.1$
     &    $1.9 \pm 0.1$ & \phn4.9 &    13.0
  \enddata
  \tablecomments{From left to right the columns give: the cluster name,
    redshift, \chandra\ RA centroid, \chandra\ dec centroid,
    \chandra-derived mass, \chandra-derived \thetafive,
    \xmm-like \thetafive,
    \planck\ MMF3 SNR, and Bolocam SNR.}
  \label{tab:sample}
\end{deluxetable*}

  \subsection{\planck}
    \label{sec:planck}

    \begin{figure*}
      \centering
      \includegraphics[height=0.25\textwidth]{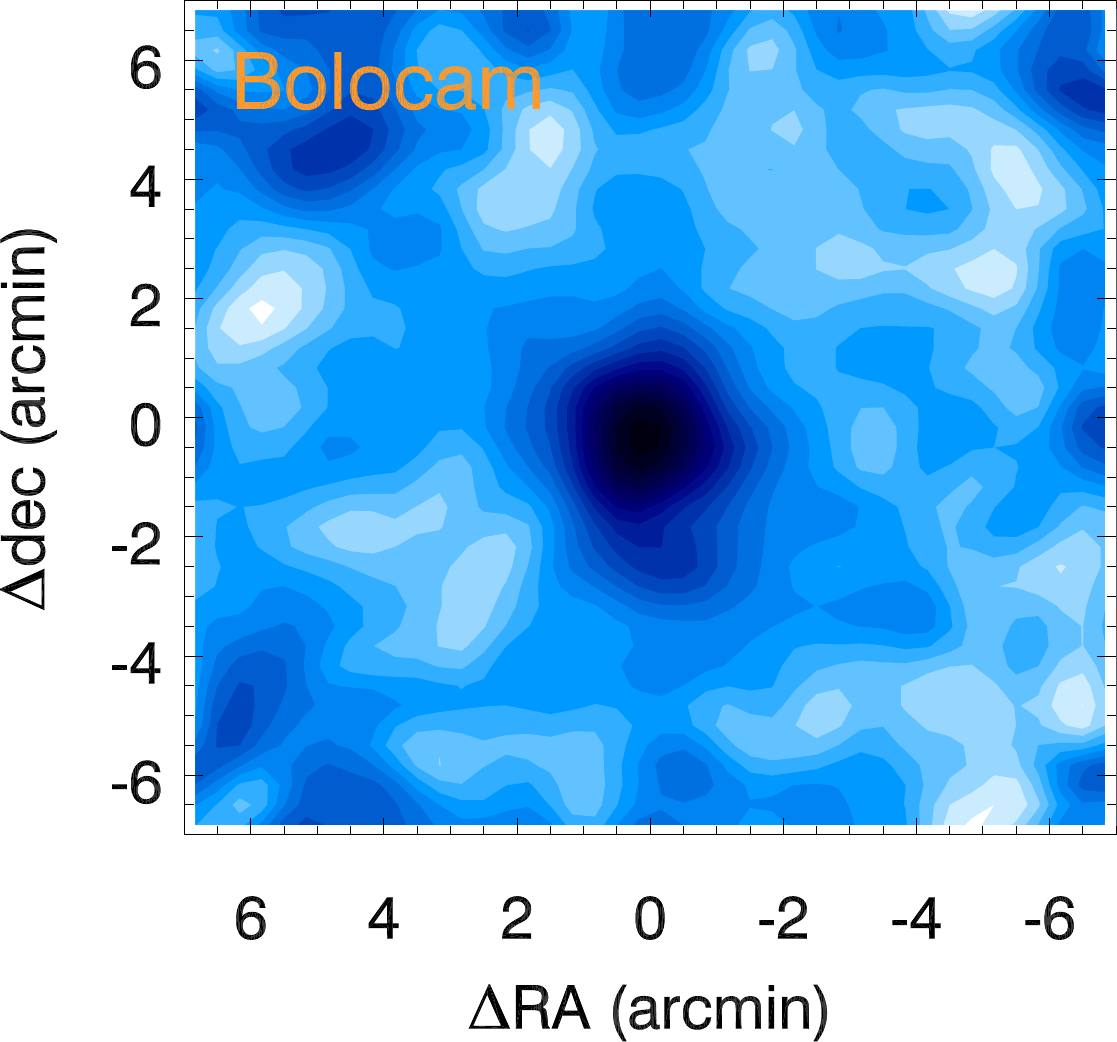}
      \hspace{0.03\textwidth}
      \includegraphics[height=0.25\textwidth]{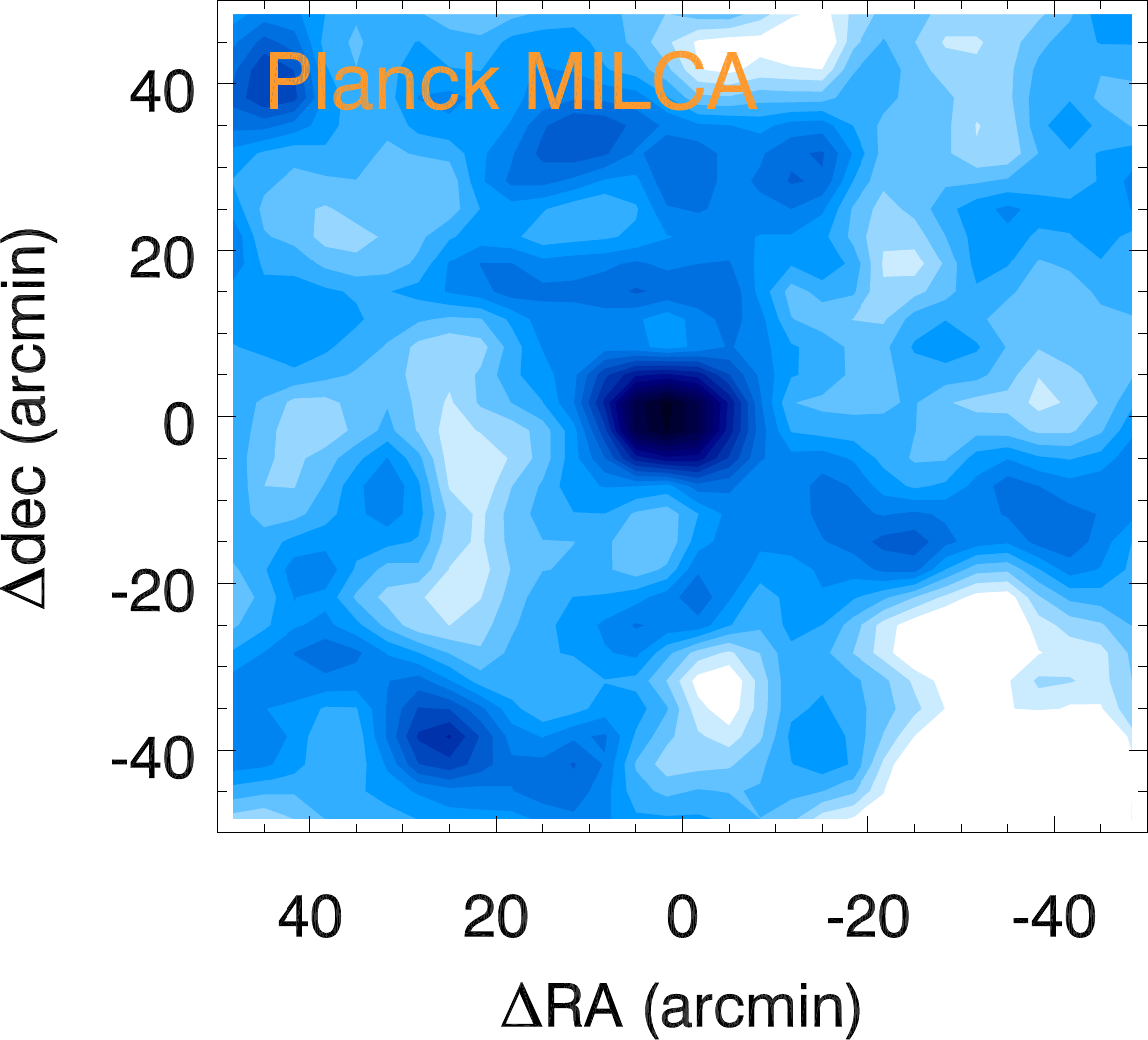}
      \hspace{0.03\textwidth}
      \includegraphics[height=0.25\textwidth]{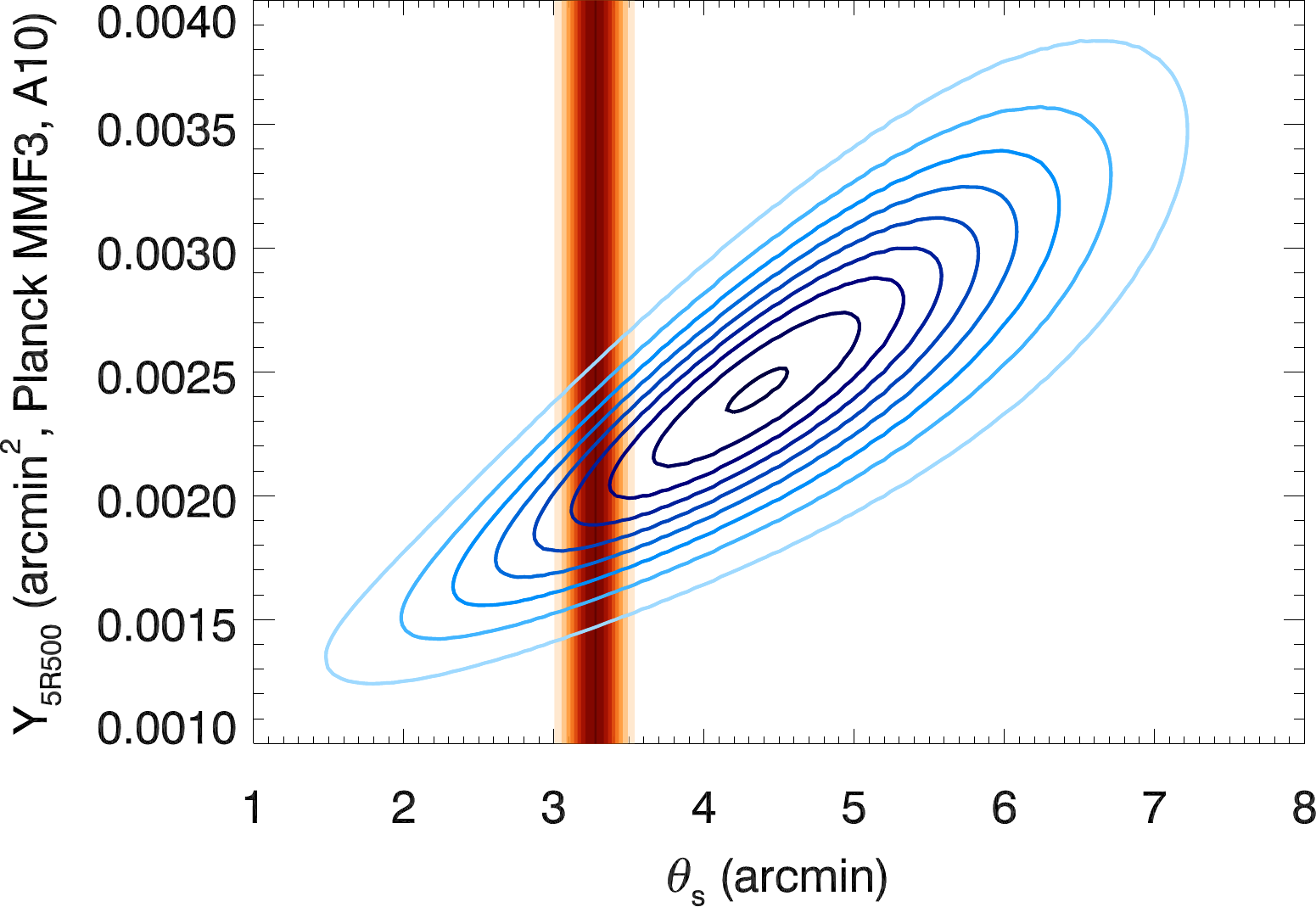}

      \vspace{0.05\textwidth}
      \includegraphics[height=0.25\textwidth]{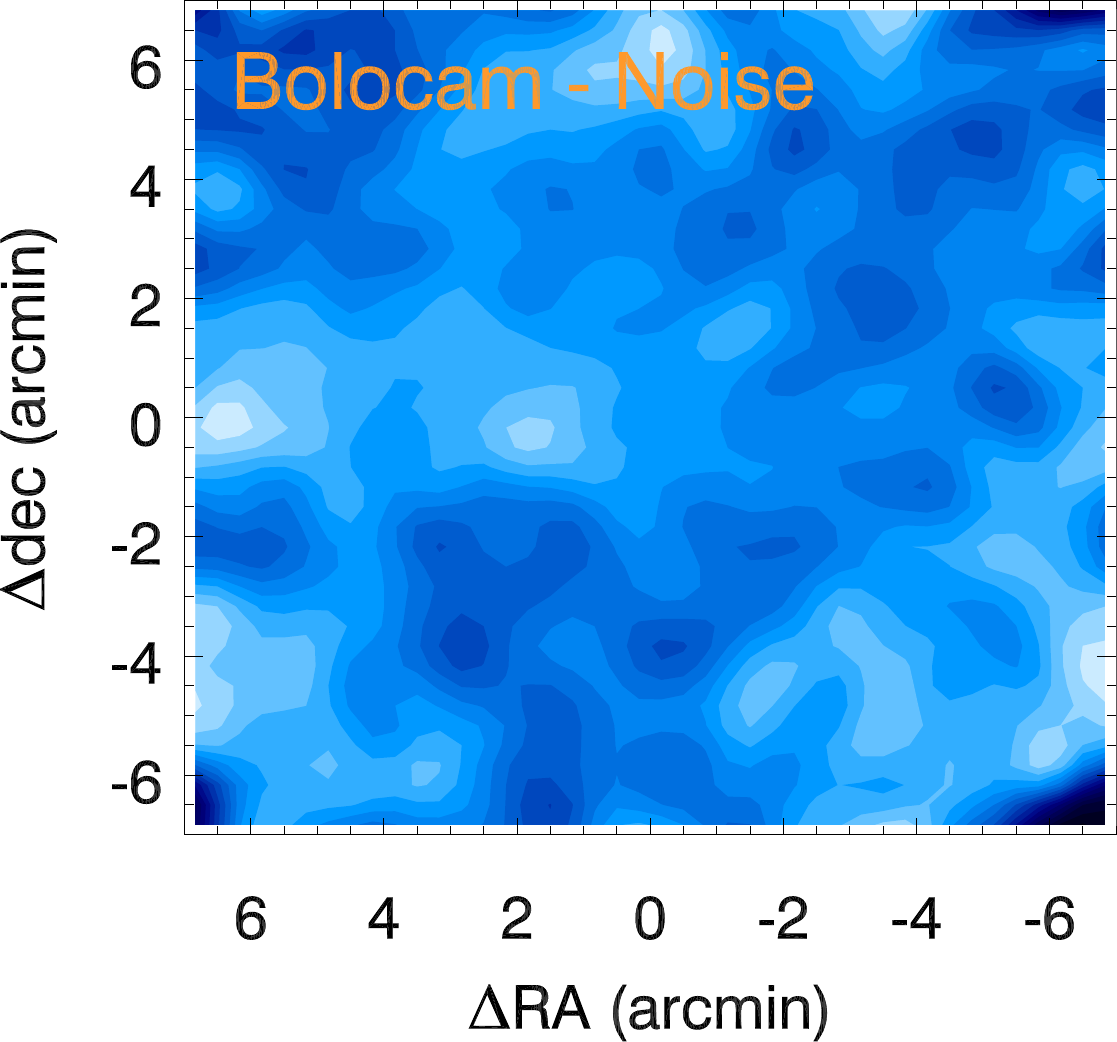}
      \hspace{0.03\textwidth}
      \includegraphics[height=0.25\textwidth]{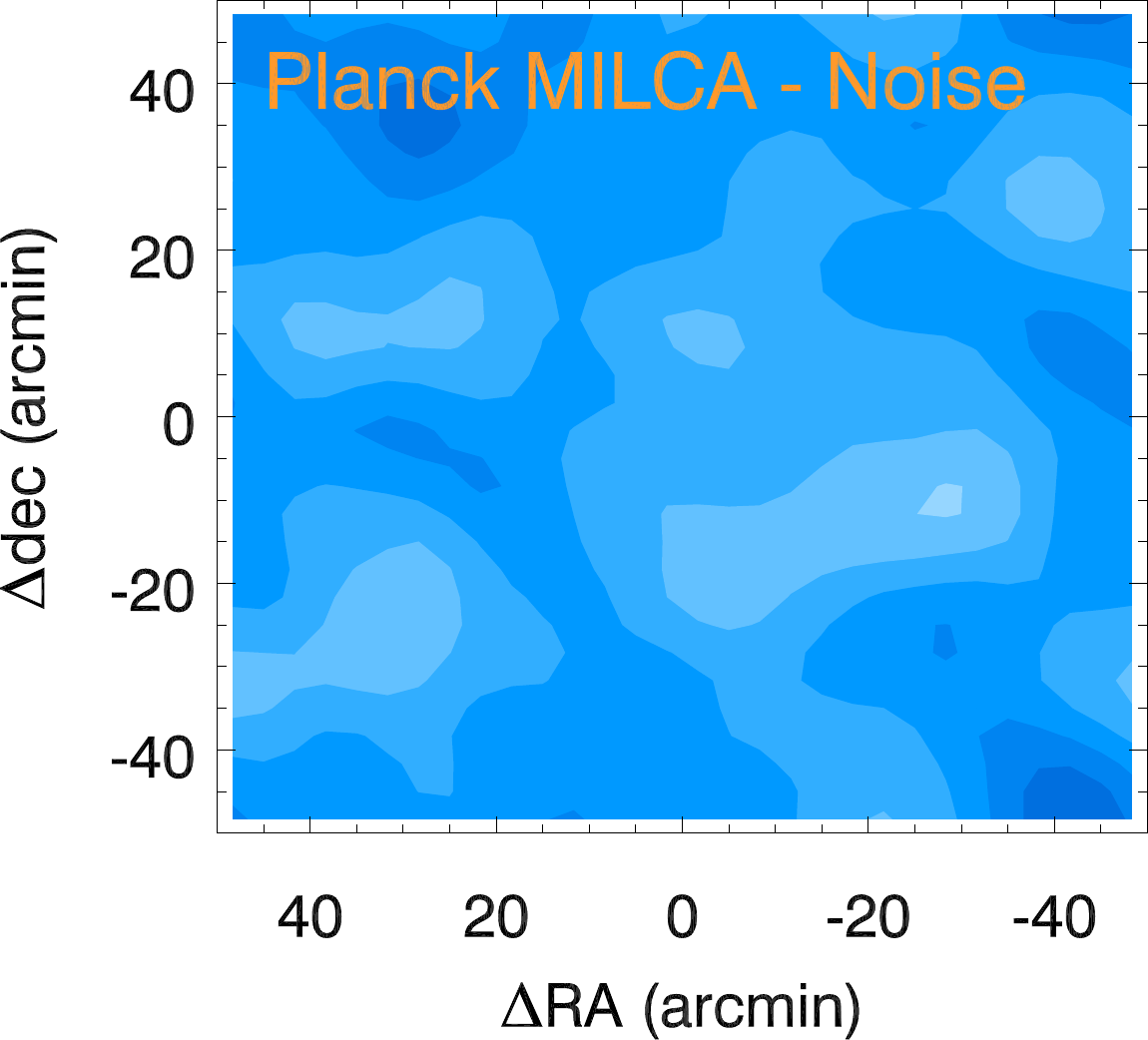}
      \hspace{0.03\textwidth}
      \includegraphics[height=0.25\textwidth]{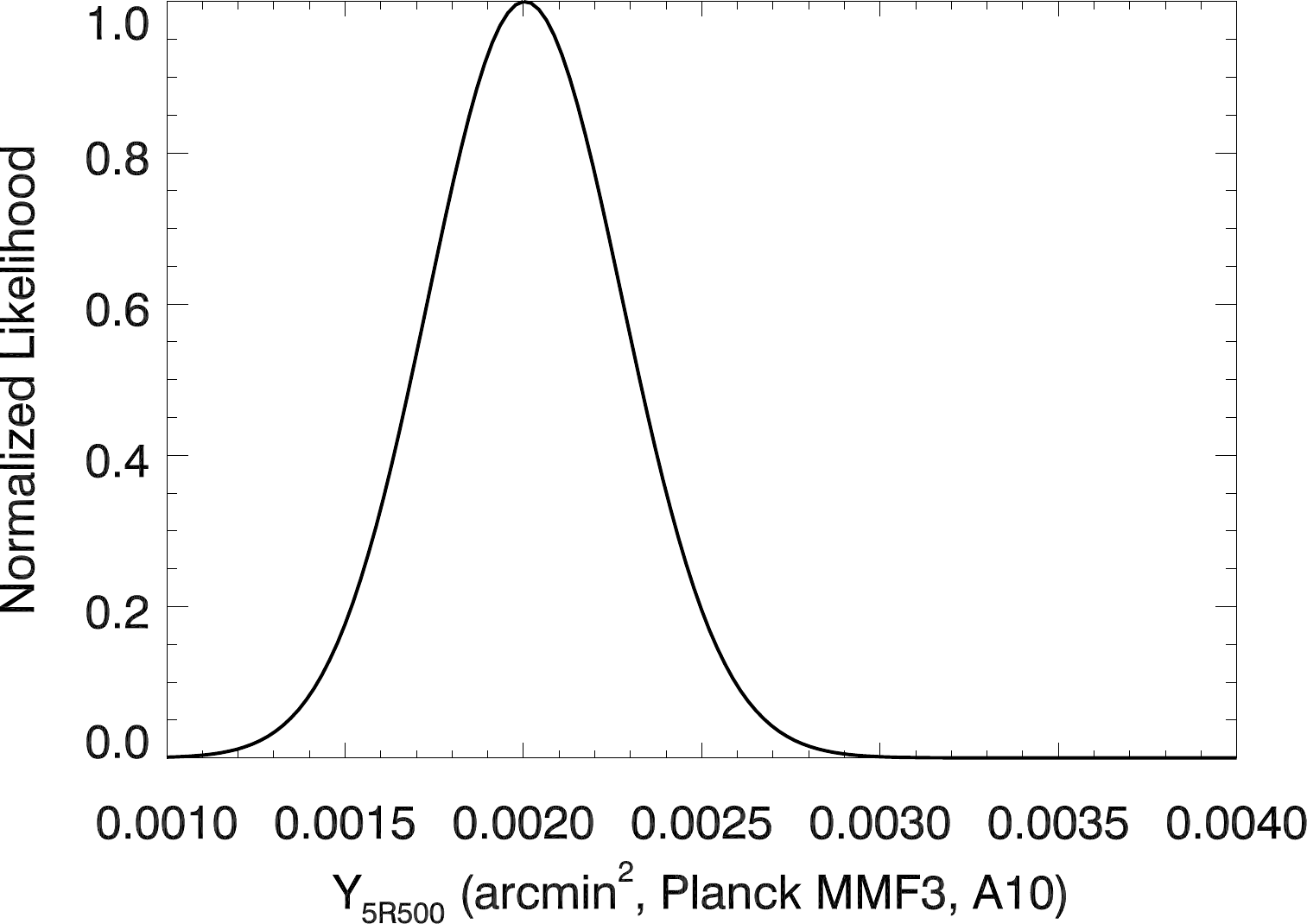}

      \caption{Examples of the SZ data used in this analysis for the cluster
        Abell 370. Left: Bolocam data (top) and 1 of the 1000 noise realizations (bottom)
        smoothed to an effective 
        FWHM of $1.4'$ for visualization (the unsmoothed data are used for
        all analyses). Middle: \planck\ MILCA \ymap\ (top) and 1 of the 1000 noise
        realizations (bottom). Right, top: \planck\ MMF3 PDF (blue contours
        separated by 0.1 in probability) and the \planck\ \xmm-like constraint on the value
        of \thetas\ (solid red, with each color separated by 0.1 in probability).
        Right, bottom: One-dimensional constraint on the value of \ybig\
        based on the \planck\ PDF and the \xmm\ prior on \thetas.}
      \label{fig:MMF3_PDF}
    \end{figure*}
    \vspace{12pt}

    The 2015 \planck\ data release\footnote{
      \url{http://irsa.ipac.caltech.edu/data/Planck/release_2/docs/}}
    contains a range of products related to the SZ signal toward
    clusters, and this analysis utilizes two of those products:
    1) the R2.08 cluster catalog created with the MMF3 detection algorithm,
    which was the baseline catalog for the \planck\ cluster
    cosmology analysis \citep{Planck2015_XXIV}, and 
    2) the R2.00 all-sky \ymap s created based on the MILCA algorithm
    \citep{Planck2015_XXII},
    which, as detailed below, show good consistency
    with the MMF3 measurements for the clusters in the \boxs\ sample.

    The MMF3 catalog provides a two-dimensional probability
    density function (PDF) for each cluster as a function of \ybig\ and \thetas\
    assuming an A10 profile. 
    A constraint on \ybig\ can therefore be obtained by marginalizing
    over \thetas, either with or without a prior.
    As an example of such a prior, the MMF3 catalog provides the values of \mfive\
    derived from the \planck\
    data, based on a scaling relation calibrated using hydrostatic
    masses from \xmm,\footnote{
      Because these masses and \thetas\ values
      are calibrated based on \xmm\ measurements,
      they are referred to throughout this manuscript as ``\xmm-like''.}
    and these values of \mfive\ provide a direct constraint
    on \thetas\ for an assumed value of \cfive\
    (see Figure~\ref{fig:MMF3_PDF}).

    In addition to the MMF3 catalog, the value of \ybig\ can also be derived
    using the all-sky MILCA \ymap\ by fitting an
    A10 model directly to the map according to the following 
    procedure. 
    First, a prior on the value of \thetas\ from the \xmm-like
    measurements is used to set the angular size of the model.
    Then, the three-dimensional model of the cluster is projected
    to a two-dimensional image with
    the line-of-sight projection extending to a radial distance of $5 \times R_{500}$. 
    Next, the model is convolved with a $10'$ full-width half-maxima (FWHM) Gaussian 
    profile to
    match the point spread function (PSF) of the MILCA \ymap,
    and binned into square pixels with sides of $3.33'$.
    To compare to this candidate model, 
    the full-sky \texttt{HEALPix}\footnote{\url{http://healpix.jpl.nasa.gov}}
    MILCA \ymap\ data are rebinned into $100' \times 100'$ thumbnails
    centered on each cluster with identical $3.33'$ square pixels
    (see Figure~\ref{fig:MMF3_PDF}). 
    Next, 1000 random noise maps are generated from the sum of the inhomogeneous
    noise map and the full-sky homogeneous noise spectrum under the assumption
    that the noise is Gaussian. From these noise realizations, a
    variance per pixel is computed, and the inverse of this variance
    is used as a weighting factor when fitting the A10 model to the data.
    The fits are performed using the generalized least squares routine
    \texttt{MPFITFUN} \citep{Markwardt2009}, and the only free parameter in the fits is the
    overall normalization of the A10 model, \ybig.

    The homogeneous noise spectrum of the MILCA \ymap\ is not white, and therefore 
    the per-pixel variance of the random noise maps does not fully describe the data.
    As a result, the weighting factors used in the fits are in general
    sub-optimal.
    This causes the derived parameter uncertainties from the fits to be
    larger than those from an optimal fit, but it does not produce any
    bias in the parameter values.
    However, the parameter uncertainties will in general be mis-estimated using
    this procedure.
    Consequently, rather than estimating these uncertainties using the per-pixel
    variance, they are determined using the 1000 noise realizations.
    Specifically, the best-fit
    model obtained from the data is added to each of the 1000 noise
    realizations, all of which are then fit using the same procedure
    as applied to the real data.
    For each of these fits, the value of \thetas\ is varied according to
    its prior, thus fully including these uncertainties.
    The spread of values obtained for a given parameter based on these 1000 fits then
    provides the uncertainty on that parameter.
    
\begin{figure*}
      \centering 
        \includegraphics[height=0.33\textwidth]{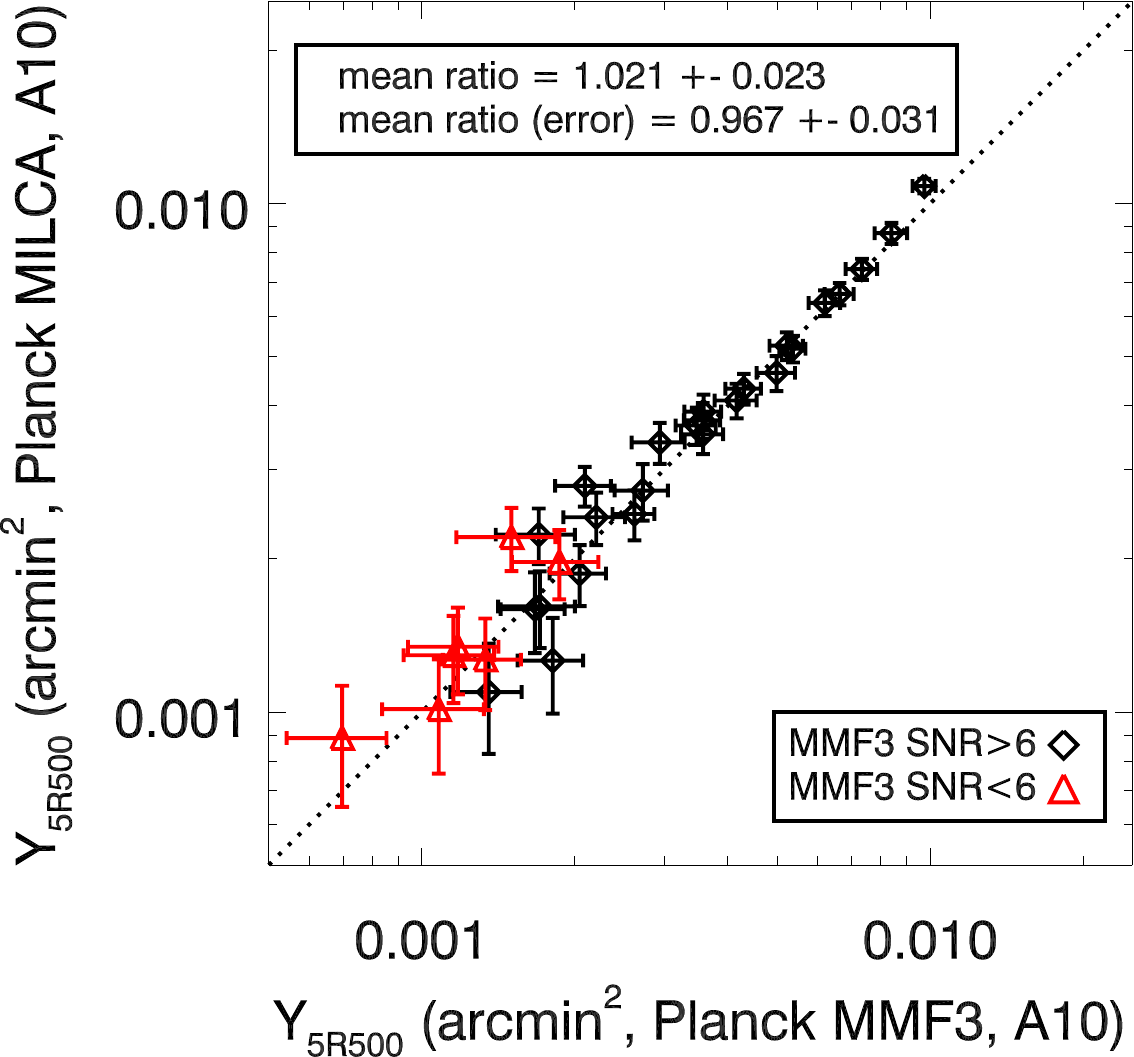}
        \hspace{0.03\textwidth}
        \includegraphics[height=0.33\textwidth]{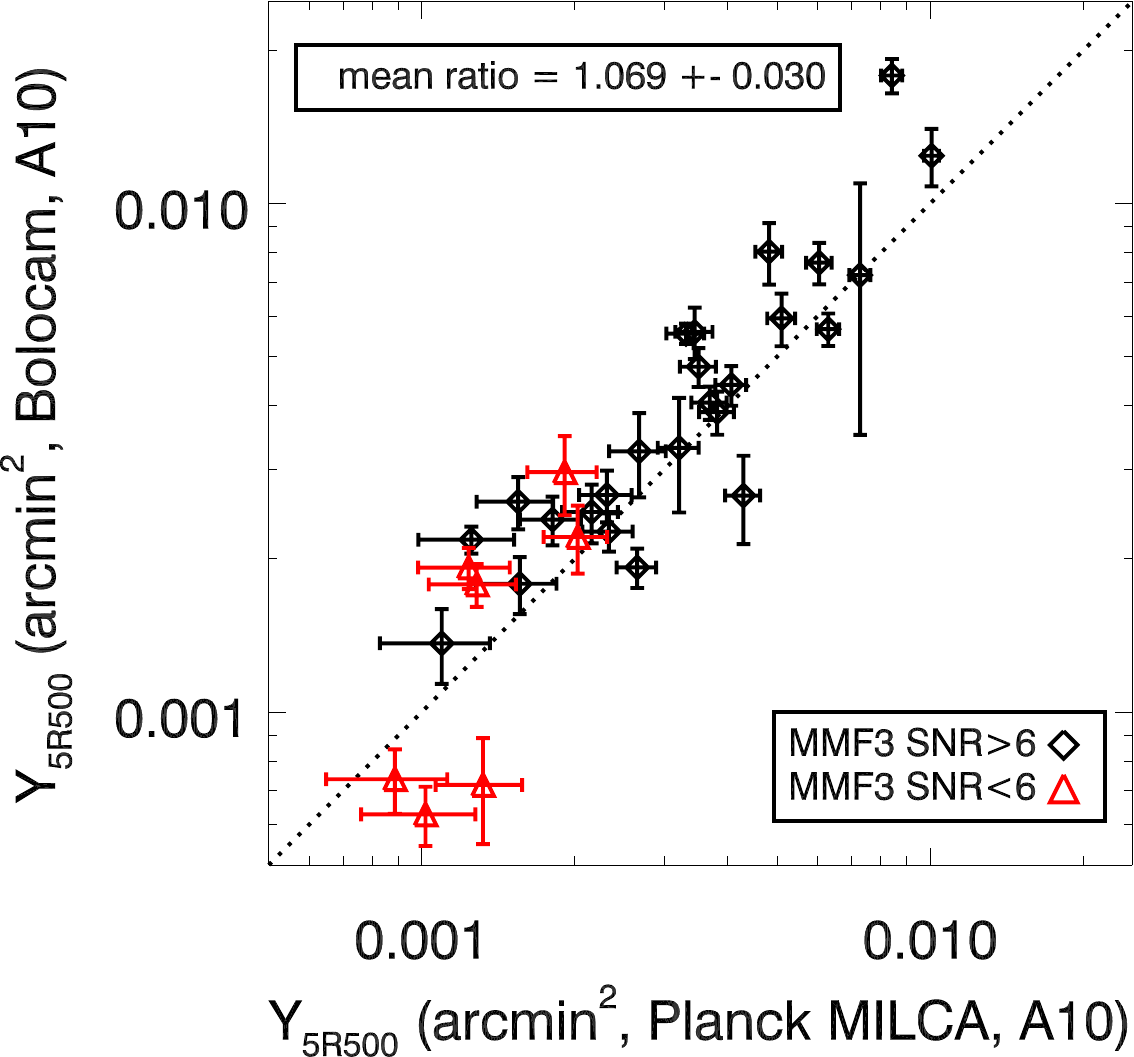}
      \caption{Left: the value of \ybig\ obtained from A10 fits to the \planck\ MILCA
        \ymap s compared to the value
        of \ybig\ recovered from \planck\ using the MMF3 algorithm. 
        On average, the two results are consistent.
        Right: the value of \ybig\ obtained from A10 fits to the Bolocam
        data compared to the value of \ybig\ recovered from the \planck\
        MILCA \ymap s.
        Given the 5\% flux calibration uncertainty on the Bolocam data,
        the two results are consistent on average.
        In both plots, clusters above the \planck\ cluster cosmology
        cut (MMF3 SNR $>6$) are shown in black, while MMF3 detections
        below the cut are shown in red.
        Both plots contain all 32 \boxs\ clusters detected by \planck\
        using the MMF3 algorithm.}
      \label{fig:a10_comparison}
    \end{figure*}

    Based on the above fits, the value of \ybig\ obtained from the
    MMF3 catalog is consistent with the value of \ybig\ obtained
    from the MILCA \ymap, with a sample-mean ratio of $1.021 \pm 0.023$
    for the 32 \boxs\ clusters contained in the MMF3 catalog
    (see Figure~\ref{fig:a10_comparison}).
    Further, the uncertainty on \ybig\ is also consistent between
    the two, with a sample-mean ratio of $0.967 \pm 0.031$.\footnote{
      An identical fitting procedure was also applied to the \planck\
      NILC \ymap s.
      While the value of \ybig\ is consistent between
      the NILC \ymap s and the MMF3 catalog with a sample-mean ratio of $0.997 \pm 0.022$,
      the recovered uncertainties from the NILC \ymap s are systematically
      lower with a sample-mean ratio of $0.834 \pm 0.026$.
      The cause of this discrepancy is not understood, and may be
      related to the fitting technique used for the \ymap s.
      As a result, the NILC \ymap s are not considered in this analysis.}
    Therefore, on average, \ybig\ values obtained from fits to the MILCA
    \ymap s are equivalent to \ybig\ values obtained from the MMF3 catalog.
    
  \subsection{Bolocam}
    \label{sec:bolocam}

    This analysis makes use of the publicly available
    filtered Bolocam maps, which contain
    an image of the cluster that has been high-pass filtered according
    to a two-dimensional transfer function included with the data.
    Analogous to the MILCA \ymap s,
    1000 noise realizations of the Bolocam maps are provided.
    The A10 model fits are performed using the same procedure applied to the
    MILCA \ymap s, with the following differences:
    1) the Bolocam data have a $58''$ FWHM PSF, 
    2) the model must be convolved with the transfer function of the spatial
    high-pass filter, and
    3) the transfer function of the mean signal level of the map is equal
    to 0, and so an additional nuisance parameter is included in the fits
    to describe the mean signal.

\section{Comparison of SZ Measurements}
  \label{sec:SZ_measurements}

    The Bolocam fit results from Section~\ref{sec:bolocam}
    can be directly compared to the \planck-derived
    results from Section~\ref{sec:planck}, which were based on identical
    A10 model shapes and \xmm-like priors on the value of \thetas,
    along with a nearly identical fitting procedure.\footnote{
      One subtlety is that the frequency-dependent relativistic corrections
      to the SZ signal were not included in any of the fits, and this
      could potentially bias the values of \ybig\ derived from \planck\
      compared to the values derived from Bolocam.
      However, this bias should be minimal for two main reasons.
      First, the most sensitive \planck\ SZ channel is centered on 143~GHz,
      which is nearly
      identical to the Bolocam observing band centered on 140~GHz.
      Second, at 140~GHz the typical relativistic corrections for the \boxs\
      clusters are $\lesssim 10$\%, and so a severe mismatch in relativistic
      correction factors would be required to significantly bias the
      comparison of \ybig\ values.}
    The weighted mean ratio between the Bolocam and \planck\ values
    of \ybig\ obtained from these fits is $1.069 \pm 0.030$.
    Given Bolocam's 5\% calibration uncertainty, which is common to
    all of the clusters and therefore acts as a 5\% uncertainty on this
    average ratio, this result indicates consistency
    (see Figure~\ref{fig:a10_comparison}).

    Other groups have also compared \planck\ SZ measurements to
    ground-based data.
    For example, \citet{Planck2013_II} fit A10 models to a set
    of 11 clusters using \xmm\ priors on \thetas\ and SZ data
    from the Arcminute Microkelvin Imager (AMI).
    They found an average ratio of $0.95 \pm 0.05$ between the
    values of \ybig\ derived from AMI and \planck, indicating good
    agreement.
    A later comparison by \citet{Perrott2015}, 
    using AMI observations of 99 clusters, found
    systematically lower values of \ybig\ from AMI relative to \planck.
    However, the value of \thetas\ was allowed to float in the fits performed
    in their analysis, and therefore some or all of the difference in \ybig\ values may
    be a result of using different pressure profile shapes when
    fitting AMI and \planck.
    More recently, \citet{Rodriguez-Gonzalvez2016} compared SZ measurements from \planck\
    and the Combined Array for Research in Millimeter-wave Astronomy 
    (CARMA-8) for a set of 19 clusters.
    Like \citet{Perrott2015}, they floated the value of \thetas\ in their fits, 
    although, unlike \citet{Perrott2015},
    they obtained consistency, with a CARMA-8/\planck\ ratio of $1.1 \pm 0.4$.
    
\section{Joint Fits to \planck\ and Bolocam and Comparisons to
  Previous Pressure Profile Results}
  \label{sec:planck_bolocam_a10}

  Motivated by the good agreement between \planck\ and Bolocam in measuring
  the value of \ybig\ based on identical A10 profile shapes, the
  data can be combined to jointly constrain a more general gNFW shape.
  Specifically, given that Bolocam and \planck\ are most sensitive to the
  gNFW shape at large radii, the value of the outer logarithmic slope $\beta$
  is allowed to vary in these fits while the other parameters are fixed
  to the A10 values.
  In order to apply these fits to the largest sample possible, namely
  the full set of 47 \boxs\ clusters, an external prior on the value
  of \thetas\ is required due to the fact that \xmm-like priors
  only exists for 32 \boxs\ clusters.
  This \thetas\ prior is obtained from previously published values
  of \mfive\ derived using data from \chandra, 
  mainly from \citet{Sayers2013_pressure} based on the
  analysis methods detailed in \citet{Mantz2010_ii}.\footnote{
    Recall from Section~\ref{sec:gnfw} that \mfive\ 
    uniquely determines \thetas\ for a given \cfive.}
  Two \boxs\ clusters are absent from \citet{Sayers2013_pressure},
  and so the \chandra-derived \mfive\ of Abell 1689 is obtained
  from \citet{Mantz2010_ii} and the \chandra-derived \mfive\ of
  Abell 2744 is obtained from \citet{Ehlert2015}.

  One subtlety is that the \chandra-derived values of 
  \mfive\ are systematically larger than the \xmm-like values.
  In particular, the \xmm-like \mfive\ values are known to be 
  biased low by $\simeq 30$\% compared to lensing masses
  \citep{vonderLinden2014, Planck2015_XXIV}, 
  while the \chandra\ \mfive\ values described above are $\simeq 10$\% higher
  than these same lensing masses \citep{Mantz2014, Applegate2016}.
  As a result, the \xmm-like values of \thetas\ are
  smaller than the \chandra\ values of \thetas, with
  an average ratio of 1.16 for the 32 \boxs\ clusters 
  in the MMF3 catalog.
  Since \thetas\ sets the angular scale of the gNFW profile, this
  is equivalent to a change in the value of \cfive.
  However, since $\beta$ is allowed to vary in these fits,
  and $\beta$ and \cfive\ are highly degenerate over the angular
  scales probed by \planck\ and Bolocam, 
  this differing choice of \thetas\ values does not significantly
  impact the derived profile shape in the radial range
  where Bolocam and \planck\ are sensitive,
  though the specific value of $\beta$ derived from these
  fits does depend on the choice of \thetas\ (i.e., of \cfive).

  \begin{figure*}
    \centering
      \includegraphics[width=0.36\textwidth]{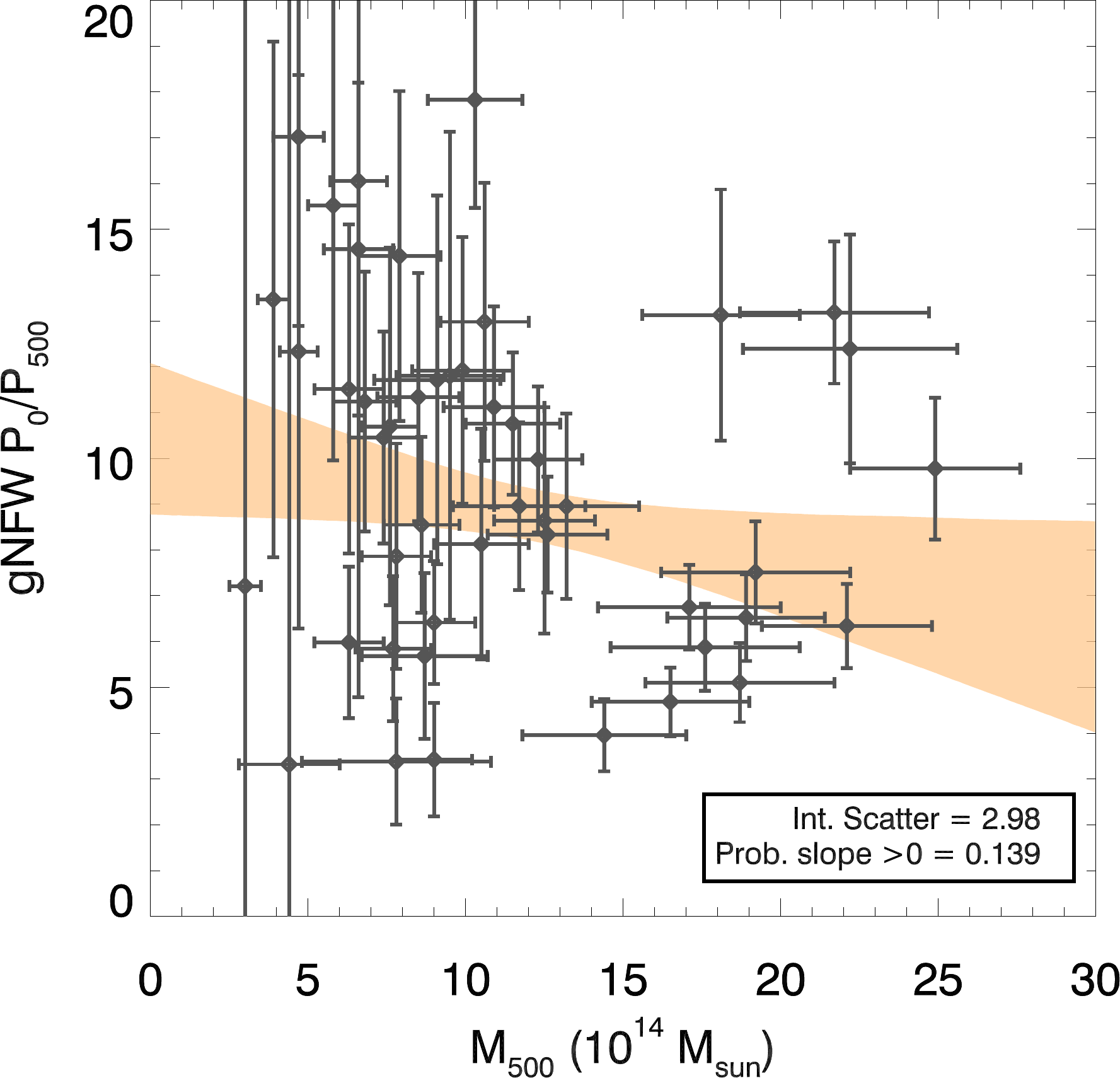}
      \hspace{0.04\textwidth}
      \includegraphics[width=0.36\textwidth]{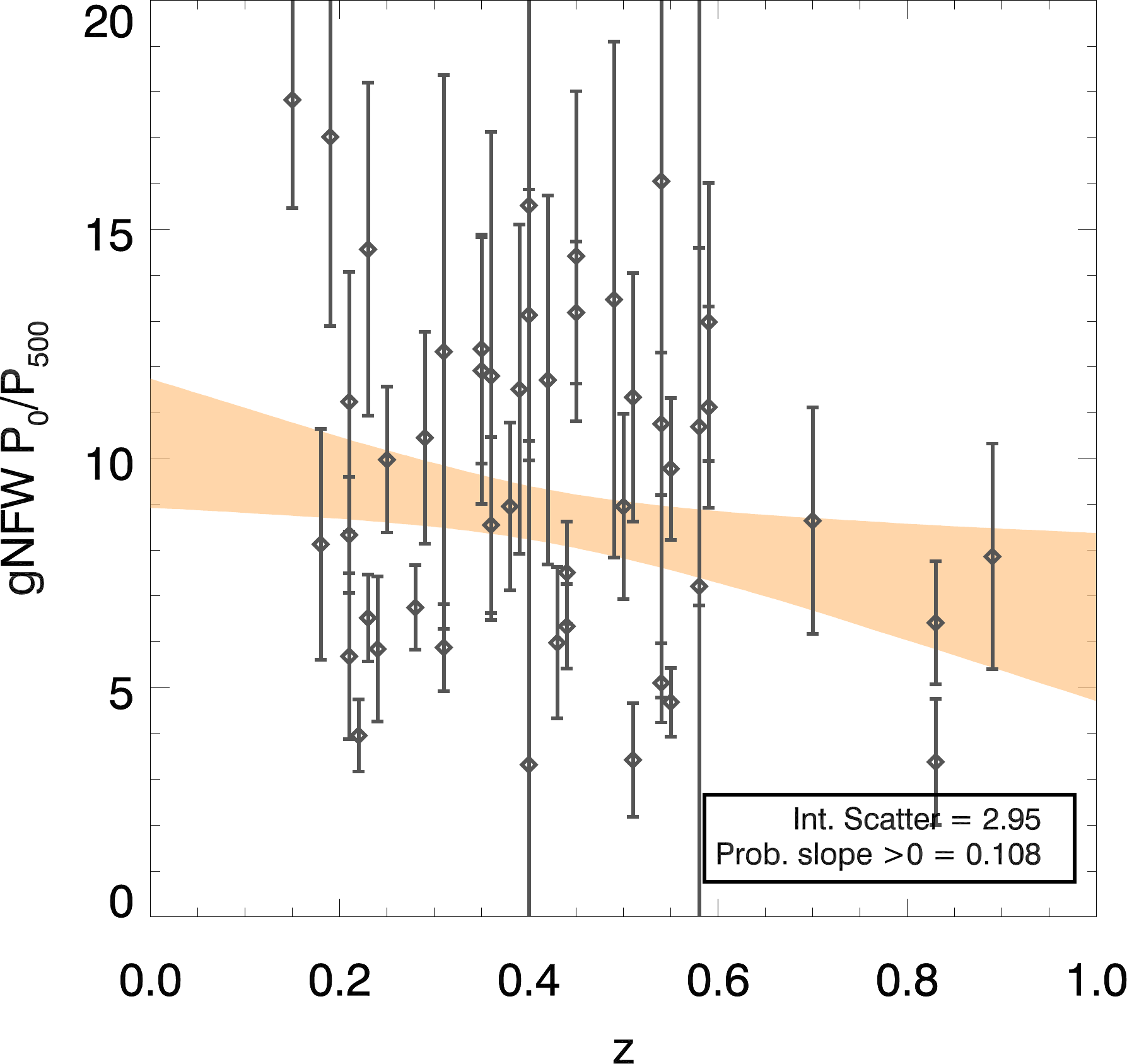}

      \vspace{6ex}

      \includegraphics[width=0.36\textwidth]{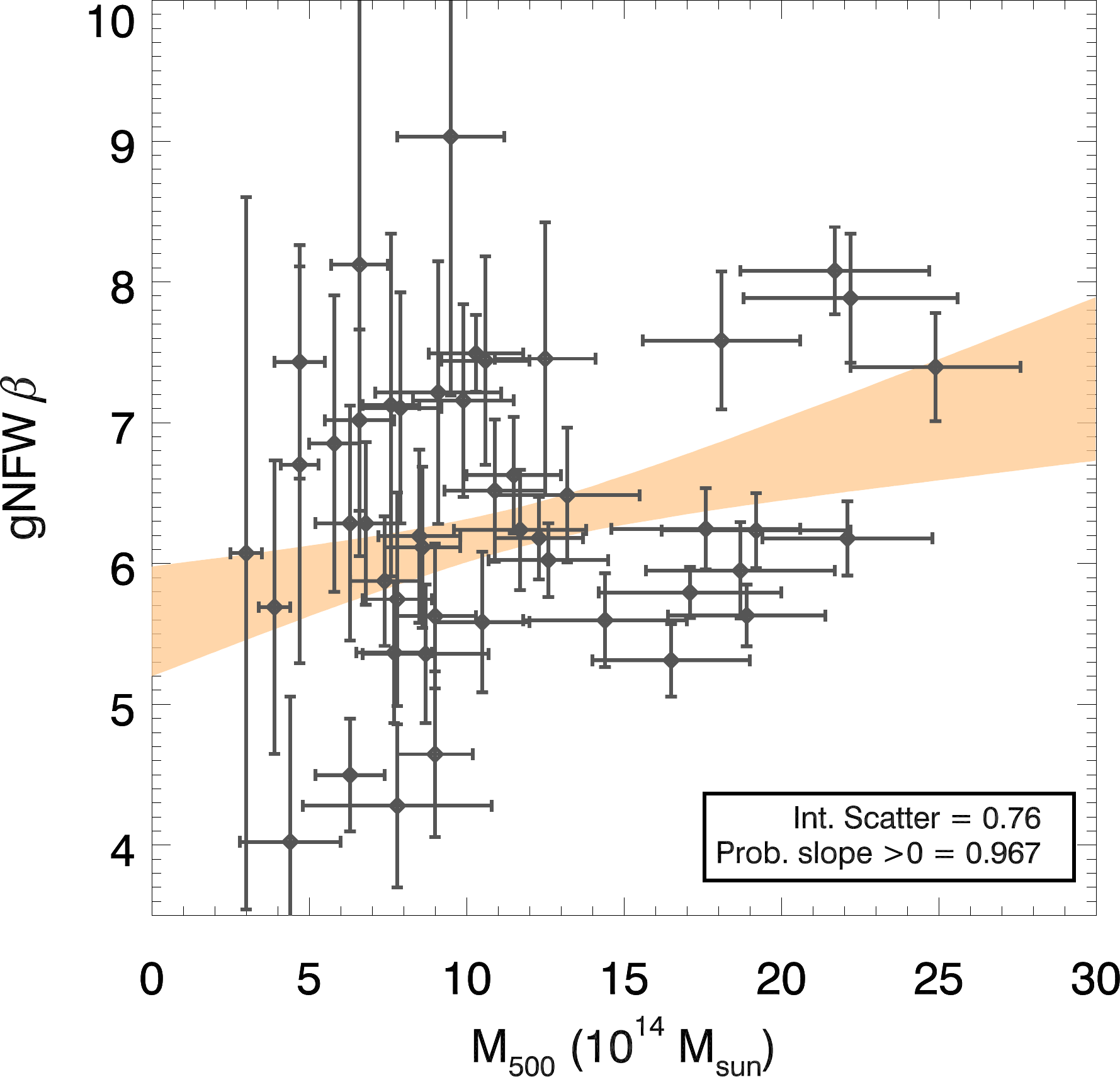}
      \hspace{0.04\textwidth}
      \includegraphics[width=0.36\textwidth]{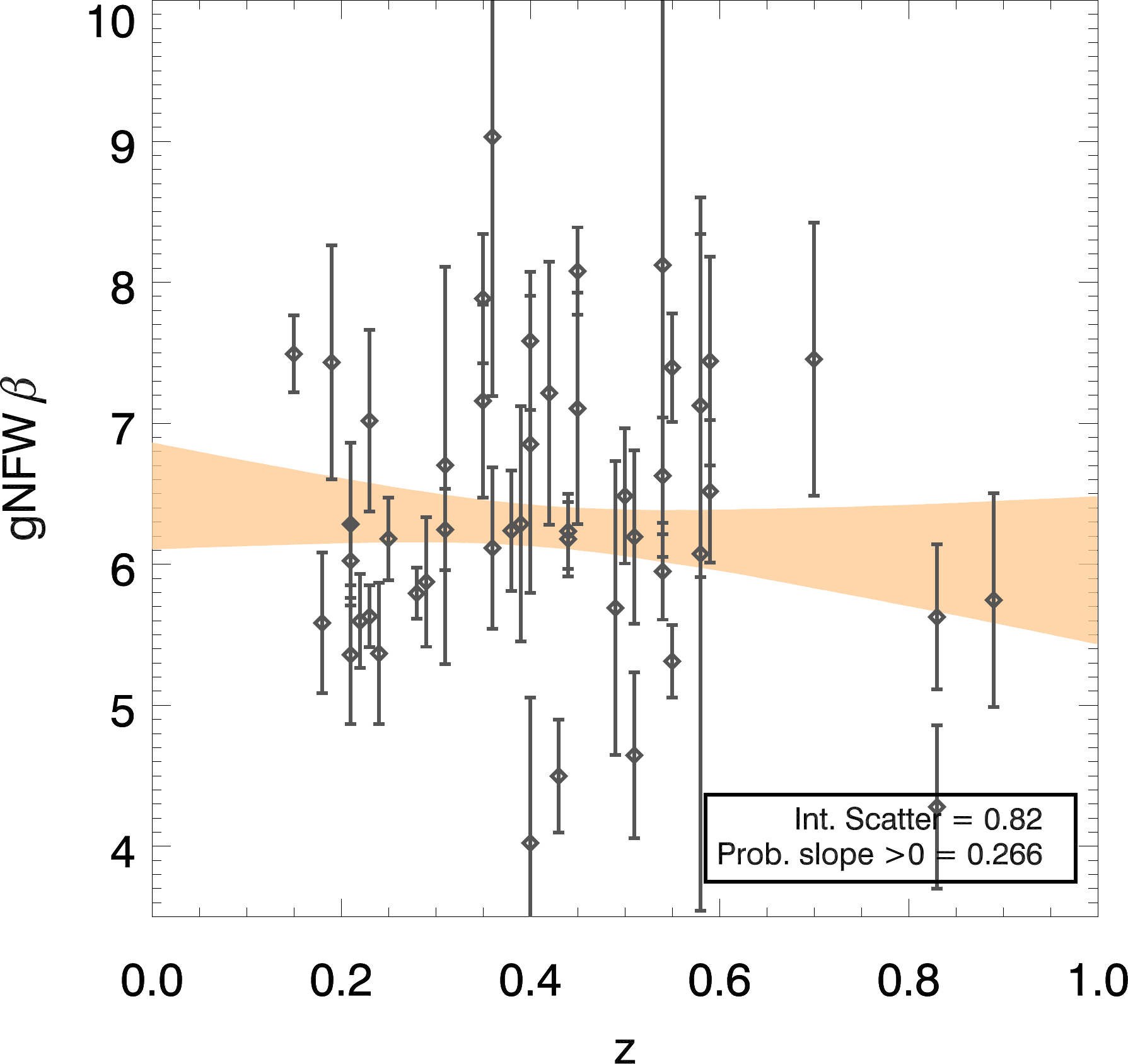}
    \caption{Best-fit parameters of the joint gNFW fit to Bolocam and \planck.
      The normalization $P_0$ (top row) and outer logarithmic slope $\beta$ 
      (bottom row) were allowed
      to float, while $C_{\textrm{500}}$, $\alpha$, and $\beta$ were fixed
      to the A10 values using a prior on \thetas\ from \chandra. 
      From left to right the plots indicate cluster mass and
      redshift, with 68\% confidence intervals of linear fits from 
      \texttt{LINMIXERR} overlaid in orange.
      At the median mass of the \boxs\ sample, the linear fits versus \mfive\ provide values of
      $P_0/P_{500} = 9.13 \pm 0.68 \pm 2.98$ and 
      $\beta = 6.13 \pm 0.16 \pm 0.76$.}
    \label{fig:linmix_gnfw}
  \end{figure*}

  To better understand the results of these jointly constrained gNFW models,
  linear fits of $P_0$ and $\beta$ were performed versus
  \mfive\ and $z$ using \texttt{LINMIXERR} \citep{Kelly2007}, 
  with the results shown in Figure~\ref{fig:linmix_gnfw}.\footnote{
    In determining $P_0$,
    relativistic corrections are applied based on spectroscopic
    \chandra\ measurements from \citet{Sayers2013_pressure}
    \citet{Mantz2010_ii}, and \citet{Babyk2012}, using
    on an effective observing frequency of 140~GHz.}
  Only modest correlations
  exist and the strongest trend is found in $\beta$ versus \mfive.
  These fits find a cluster-to-cluster scatter of $\simeq 30$\% for the value of 
  $P_0$ and $\simeq 15$\% for the value of $\beta$.
  If the linear fits versus mass are evaluated at the median value for the \boxs\ sample,
  $M_{500} = 9.5 \times 10^{14}$~M$_{\sun}$,
  then the results are $P_0/P_{500} = 9.13 \pm 0.68 \pm 2.98$ and 
  $\beta = 6.13 \pm 0.16 \pm 0.76$
  (where the first value represents measurement uncertainty and the second
  indicates intrinsic cluster-to-cluster scatter).
  Compared to the A10 model, with $P_0/P_{500} = 8.40$ and $\beta = 5.49$,
  both of these values are slightly larger and indicate a higher pressure
  in the cluster center with a steeper fall-off at large radius.
  However, in interpreting these results, it is important to note that,
  while $\beta$ provides one metric for understanding the pressure profile
  at large radius, it does not uniquely describe a single shape due
  to the strong degeneracies between the gNFW parameters.
  A more robust metric is the ratio between the integrated SZ signal
  at \rfive\ and at $5 \times R_{500}$,\footnote{
    While \yfive/\ybig\ is a more robust metric than $\beta$ for
    comparing outer profile shapes, the general convention in
    the literature has been to report gNFW fit parameters directly.
    Therefore, the comparisons presented in this section 
    generally include both values.}
  with \citet{Arnaud2010} obtaining \yfive/\ybig\ $=0.56$.
  This result can be compared to the value of
  \yfive/\ybig $=0.66 \pm 0.02 \pm 0.10$ obtained from our joint
  Bolocam/\planck\ fits to the \boxs\ clusters
  (see Table~\ref{tab:beta}).
  
    \begin{deluxetable*}{cccc}
      \tablewidth{0pt}
      \tablecaption{gNFW Outer Profile Shapes}
      \tablehead{\colhead{Analysis} & \colhead{Data Type} & \colhead{$\beta$} & \colhead{\yfive/\ybig}}
      \startdata
        this work & SZ observations & 6.13 & 0.66 \\
        \citet{LeBrun2015} & simulations & 4.63 & 0.63 \\
        \citet{Ramos-Ceja2015} & SZ power spectrum & 6.35 & 0.69 \\
        \citet{Sayers2013_pressure} & SZ observations & 3.67 & 0.28 \\
        \citet{Planck2013_V} & SZ/X-ray observations & 4.13 & 0.48 \\
        \citet{Battaglia2012} & simulations & 5.75 & 0.63 \\
        \citet{Plagge2010} & SZ observations & 5.5\phn & 0.53 \\
        \citet{Arnaud2010} & X-ray observations/simulations & 5.49 & 0.56 \\
        \citet{Nagai2007} & X-ray observations/simulations & 5.0\phn & 0.52 
      \enddata
      \tablecomments{Measurements of the outer pressure profile shape
        in large samples of clusters. The columns show the reference to the analysis,
        the type of data used in the analysis, the value of $\beta$, and the
        value of \yfive/\ybig.
        In the case of \citet{LeBrun2015} their ``median AGN 8.0'' fits were used,
        and were scaled to the median mass of the \boxs\ sample using their
        fitting formulae.
        In the case of \citet{Battaglia2012}, their ``AGN Feedback $\Delta = 500$''
        fits were used, and were scaled to the median mass and redshift of the \boxs\ sample using their
        fitting formulae.
        Uncertainties are not available for most analyses, and so they have been omitted.}
      \label{tab:beta}
    \end{deluxetable*}

    As mentioned in Section~\ref{sec:gnfw}, a range of other analyses
    beyond \citet{Arnaud2010} have also constrained gNFW profiles
    in large samples of clusters. In particular, several groups
    have examined these profiles at large radius using either
    simulations or SZ observations.
    For example, recent simulations from
    both \citet{Kay2012} and \citet{Battaglia2012}
    note a trend of increasing $\beta$ with redshift,
    and both \citet{Battaglia2012} and \citet{LeBrun2015}
    find increasing values of $\beta$ with increasing mass.
    Specifically, evaluating the \citet{LeBrun2015} fits
    at the median mass of the \boxs\ sample yields $\beta = 4.63$
    and \yfive/\ybig\ $=0.63$, the latter indicating an outer profile
    shape that is consistent with our joint Bolocam/\planck\ fit.
    \citet{Battaglia2012} used a parameterization allowing $P_0$, $\beta$, and \cfive\
    to vary with mass and redshift according to functional forms described by, 
    for example
    \begin{displaymath}
      \beta \propto M^{b_{\textrm{M}}} (1 + z)^{b_{\textrm{z}}}.
    \end{displaymath}
    Evaluating their ``AGN Feedback $\Delta = 500$'' fit at the 
    median mass and redshift of the \boxs\ sample results in a value of
    $\beta = 5.75$ and \yfive/\ybig\ $=0.63$, both in relatively good
    agreement with our joint Bolocam/\planck\ fits.

    \begin{figure}
      \centering
      \includegraphics[width = \columnwidth]{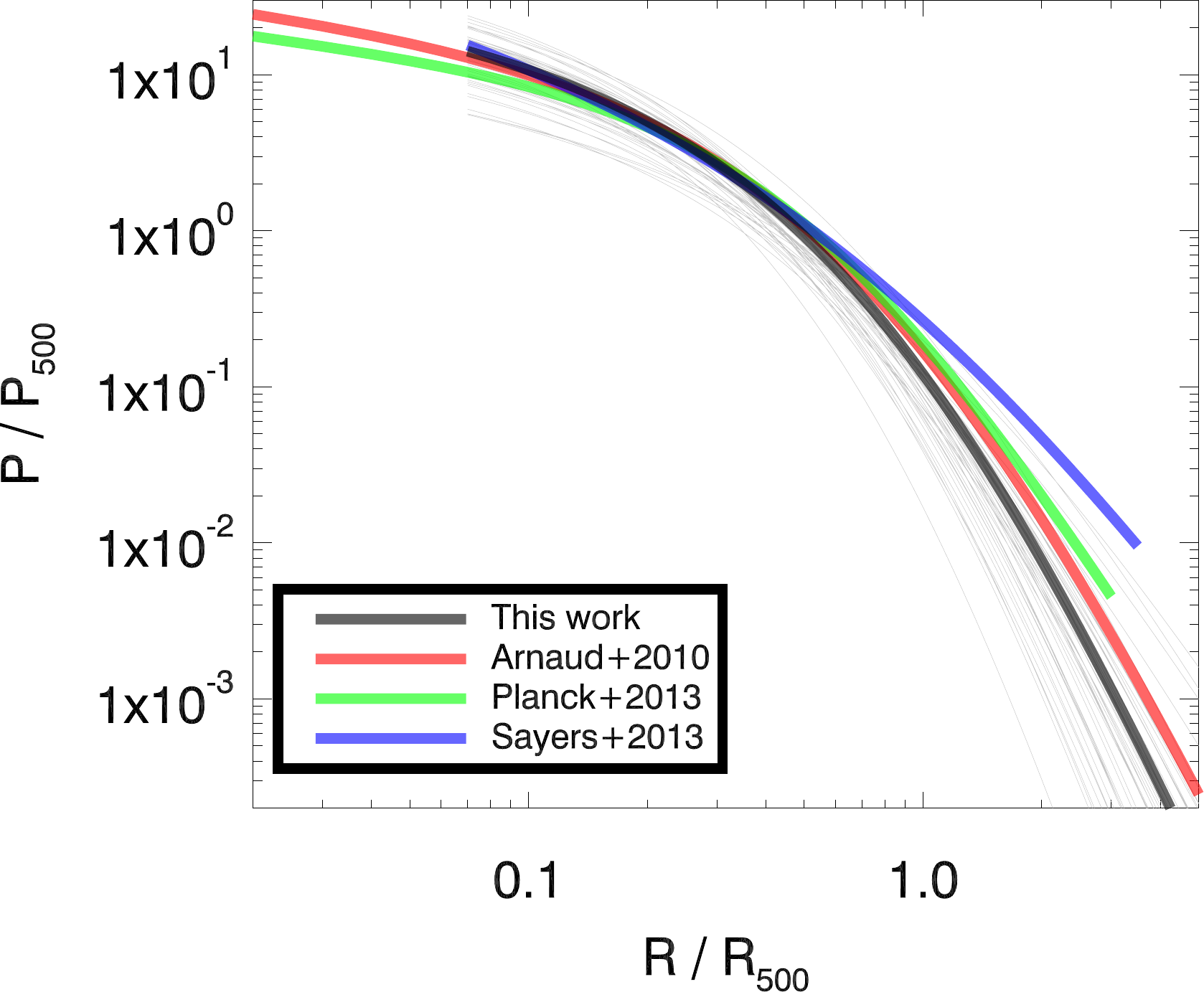}
      \caption{The ensemble-average best-fit gNFW profile to the joint Bolocam/\planck\ data
        for the \boxs\ sample of clusters (black). Profiles for the 47 individual \boxs\
        clusters are shown as thin gray lines. For comparison, the ensemble-average profiles
        from other published gNFW fits to large samples of clusters are shown
        in red \citep{Arnaud2010}, green \citep{Planck2013_V}, and blue
        \citep{Sayers2013_pressure}. The profiles extend over the approximate
        radial range probed by the data in each analysis. The ensemble-average profiles
        show good agreement at $R \lesssim 0.5$\rfive, but noticeably diverge
        at larger radii.}
      \label{fig:gnfw_profile}
    \end{figure}

    Given the good agreement of our results with \citet{Battaglia2012},
    we also fit an identical functional form to the joint
    Bolocam/\planck\ constraints on $\beta$, finding
    $b_{\textrm{M}} = 0.077 \pm 0.026$ and $b_{\textrm{z}} = -0.06 \pm 0.09$.
    These can be compared to the values of $b_{\textrm{M}} = 0.048$
    and $b_{\textrm{z}} = 0.615$ obtained by \citet{Battaglia2012}, although
    some caution is required because the values of \cfive\ were not varied in
    our fits as they were by \citet{Battaglia2012}.
    These results indicate that the trend in mass seen in the \citet{Battaglia2012}
    simulations is reproduced in our fits, but the trend in redshift is not.

    The lack of a redshift trend could be a result of the \boxs\
    sample selection, which is biased toward relaxed cool-core systems at low-$z$
    and toward disturbed merging systems at high-$z$ \citep[see][]{Sayers2013_pressure}.
    For example, 13 \boxs\ clusters are defined as relaxed based on the SPA
    criteria of \citet{Mantz2015}, and these clusters produce
    a value of $\beta = 6.83 \pm 0.37$.
    In contrast, 10 \boxs\ clusters are defined as merging based on either failing 
    the Symmetry/Alignment criteria\footnote{
      \citet{Mantz2015} use SPA to stand for symmetry, peakiness, and alignment,
      and relaxed clusters must pass a threshold in all three criteria. 
      Some known merging clusters
      pass the peakiness test, and so therefore merging clusters were
      partially selected based on failing the Symmetry and Alignment portions of the test.}
    or containing a radio relic/halo based on the
    analysis of \citet{Feretti2012} and \citet{Cassano2013}, and these
    clusters produce a value of $\beta = 5.59 \pm 0.61$.
    Therefore, an excess of cool-core clusters at low-$z$ (which have larger
    values of $\beta$ on average), and an excess of merging clusters
    at high-$z$ (which have smaller values of $\beta$ on average),
    will artificially introduce a trend of decreasing $\beta$ with
    redshift for the \boxs\ sample.

    Other groups have used SZ observations to constrain gNFW
    profile shapes at large radii.
    For example, \citet{Plagge2010} fit SZ data from a set of 15
    clusters, finding $\beta = 5.5$ and \yfive/\ybig\ $=0.53$.
    More recently, \citet{Planck2013_V} used \planck\ observations
    of a larger cluster sample to constrain $\beta = 4.13$
    and \yfive/\ybig $=0.48$
    (see Table~\ref{tab:beta} and Figure~\ref{fig:gnfw_profile}).
    Both of these analyses indicate a shallower outer profile
    than our joint Bolocam/\planck\ fits, although some
    of this difference may be a result of sample selection.
    Specifically, the \citet{Plagge2010} sample contains clusters
    with a median redshift of 0.28 and a median mass
    of $M_{500} \sim 8 \times 10^{14}$~M$_{\sun}$, and the \citet{Planck2013_V}
    sample contains clusters with a median redshift of 0.15
    and a median mass of $M_{500} = 6.3 \times 10^{14}$~M$_{\sun}$.
    If the parameterization of \citet{Battaglia2012}
    is used to rescale their gNFW fits to the median
    mass and redshift of the \boxs\ sample, then the
    resulting value of \yfive/\ybig\ 
    from both the \citet{Plagge2010} and the \citet{Planck2013_V} fits is equal
    to 0.56, closer to our value of $0.66 \pm 0.02 \pm 0.10$.
    The cause of the remaining difference is unclear, although
    it could be related to the mass estimates used in these
    analyses. In particular, \citet{Planck2013_V} used \xmm-derived
    masses to set the value of \thetafive, and, as noted above, these masses are
    known to be biased low, resulting in a different profile shape,
    and thus \yfive/\ybig\ ratio, 
    for a given set of gNFW parameters. 

    In another recent work, \citet{Sayers2013_pressure} obtained,
    from a joint fit to Bolocam observations of all the clusters
    in the BoXSZ sample, $\beta = 3.67$ and \yfive/\ybig\ $=0.28$,
    with an overall profile that noticeably diverges from our
    joint Bolocam/\planck\ fit at large radius.
    This is particularly surprising because
    the cluster samples are nearly identical, and the only significant 
    difference is the inclusion of \planck\ data in our current analysis.
    Because the Bolocam observations were made from the ground at a single
    observing frequency, they have
    have reduced sensitivity to large angular scales as a result of both
    atmospheric fluctuations and primary CMB anisotropies.
    In contrast, \planck\ is able to remove CMB anisotropies via its multiple
    observing channels, and it is not subject to atmospheric fluctuations.
    Therefore, the \planck\ data are likely to provide more
    robust constraints on large angular scales.
    Though efforts were made in \citet{Sayers2013_pressure} to account
    for the atmospheric and CMB noise, they may be the primary cause
    of the shallower outer profile found in that work.

    Beyond these SZ observations of large samples of individual clusters,
    \citet{Ramos-Ceja2015} used measurements of the SZ power spectrum
    on small angular scales from the South Pole Telescope \citep[SPT,][]{Reichardt2012}
    to constrain the average pressure profile shape.
    They found that the A10 model needs to be adjusted to have an
    outer slope $\beta = 6.35 \pm 0.19$ (\yfive/\ybig $ = 0.69 \pm 0.03$)
    in order to match the SPT measurements.
    Further, if this value of $\beta$ is adopted, then their
    analysis implies little or no evolution in its value
    as a function of redshift.
    These results are consistent with our findings.

\section{Test of the \planck\ Cluster Completeness Estimate}
  \label{sec:implications}

    \begin{figure}
      \centering \includegraphics[width = 0.36\textwidth]{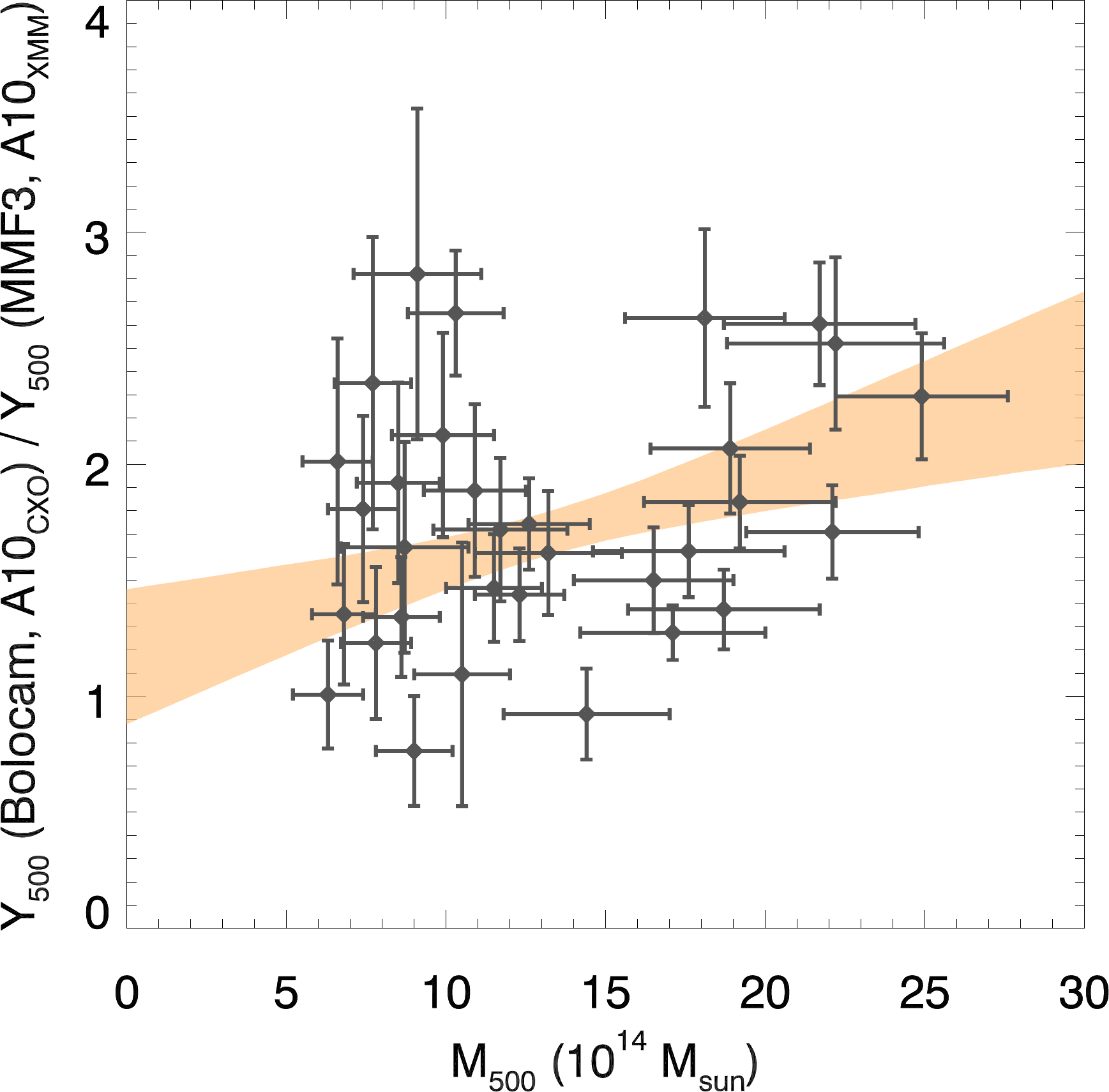}
      \caption{The ratio of \yfive\ measured from A10 fits to
        Bolocam using a \chandra\ prior on \thetas\ and \yfive\
        measured from the \planck\ MMF3 algorithm using the 
        A10 model with an \xmm-like prior on \thetas.
        The \chandra\ value of \thetas\ is larger by an average
        fraction of 1.16, resulting in systematically larger values
        of \yfive\ measured by Bolocam.
        The \yfive\ ratio is shown as a function of \mfive, with the 68\%
        confidence interval of linear fits from \texttt{LINMIXERR}
        overlaid in orange.
        This linear fit provides a mapping from the Bolocam measurements
        of \yfive\ to the \planck\ measurements of \yfive, allowing for a test
        of the \planck\ completeness using the Bolocam data.}
      \label{fig:linmix_m500}
    \end{figure}

    An accurate characterization of the completeness of the \planck\ cluster
    survey is required for cosmological analyses, and the discrepancy
    between the Planck cluster and CMB power spectrum cosmological results
    motivates special attention to such a characterization \citep{Planck2015_XXIV}.
    The details of how
    the completeness is estimated are given in \citet{Planck2015_XXVII}
    and summarized below.
    First, a set of clusters based on spherical profiles obtained
    from simulated clusters \citep{LeBrun2014, McCarthy2014} are inserted into
    both real and simulated \planck\ maps.
    The MMF3 algorithm is then applied to these maps, and the probability
    of detecting a cluster above a given SNR
    is determined as a function
    of \yfive\ and \thetas\ based on a brute force Monte-Carlo,
    which has been publicly released as part of the MMF3 catalog.
    Ideally, the accuracy of the completeness function would be tested
    using a catalog of real clusters with known positions, \thetas, and \yfive.
    In the absence of such a catalog, \citet{Planck2015_XXVII} 
    undertook a somewhat less demanding test using the MCXC
    \citep{Piffaretti2011} and SPT \citep{Bleem2015} cluster
    catalogs, which contain cluster positions and \thetas\ values,
    but not \yfive\ values.

    The \boxs\ sample enables a better approximation of
    the ideal test of the \planck\ completeness because it has positions,
    \thetas, and \yfive\ estimates for each cluster.
    Specifically, the positions and \thetas\ values are obtained from \chandra,
    the latter rescaled by a factor of 1.16 to account for the average difference
    between the \chandra\ and \xmm\ values.
    This rescaling is required because \xmm-derived \thetas\ values were
    used to calibrate the \planck\ completeness.
    Although it would be better to use the \xmm\ \thetas\ values 
    for all of the \boxs\ clusters,
    they only exist for the clusters detected by the MMF3 algorithm,
    significantly limiting the value of such a test.
    In order to obtain \yfive\ estimates from Bolocam, the following
    procedure is applied.
    First, the Bolocam value of \yfive\ for each \boxs\ cluster
    is generated from A10 model fits to
    the Bolocam data using the \chandra\ value of \thetas.
    Next, for the 32 \boxs\ clusters in the MMF3 catalog,
    the \planck\ value of \yfive\ is derived
    from the MMF3 PDF using the \xmm-like value of \thetas\
    in order to mimic the computation of \yfive\ 
    values used in the \planck\ completeness estimate.
    The ratio of the Bolocam and \planck\ 
    \yfive\ values is then fit as a function of \mfive\
    using \texttt{LINMIXERR} (see Figure~\ref{fig:linmix_m500}).
    The results of this linear fit, including the $\simeq 25$\%
    intrinsic scatter, are then used to rescale the
    Bolocam \yfive\ measurements for all of the \boxs\ clusters.
    By fitting versus \mfive, this ensures that the mass dependence
    of the profile shape found in Section~\ref{sec:planck_bolocam_a10} is
    fully included in the conversion from Bolocam to \planck\
    measurements of \yfive.
    As part of this rescaling, an additional 5\% uncertainty is
    added to account for the Bolocam flux calibration uncertainty,
    although the overall error budget is dominated by the 
    intrinsic scatter in the linear fit.

\begin{figure*}
      \centering
      \includegraphics[width = 0.4\textwidth]{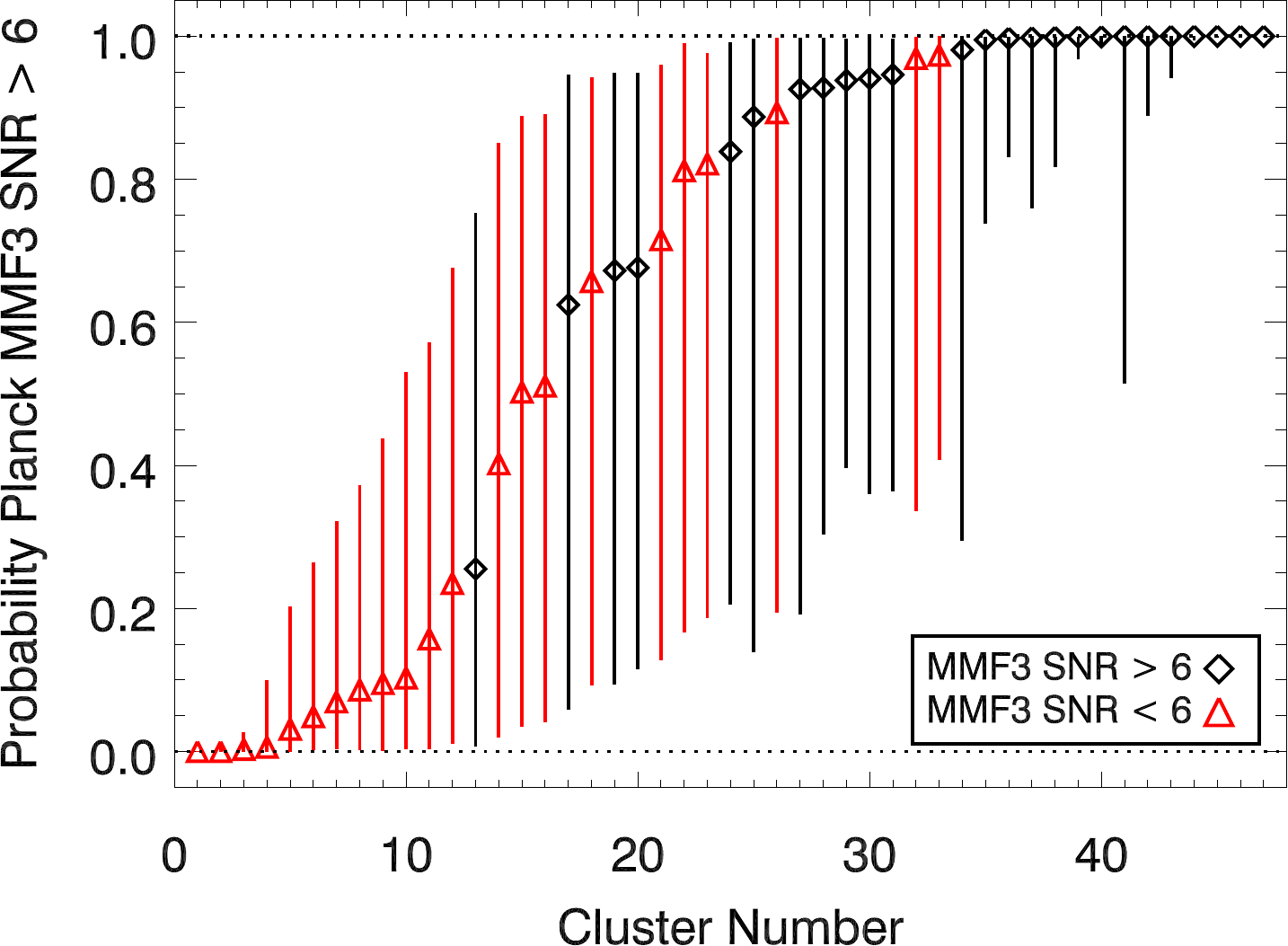}
      \hspace{0.05\textwidth}
      \includegraphics[width = 0.4\textwidth]{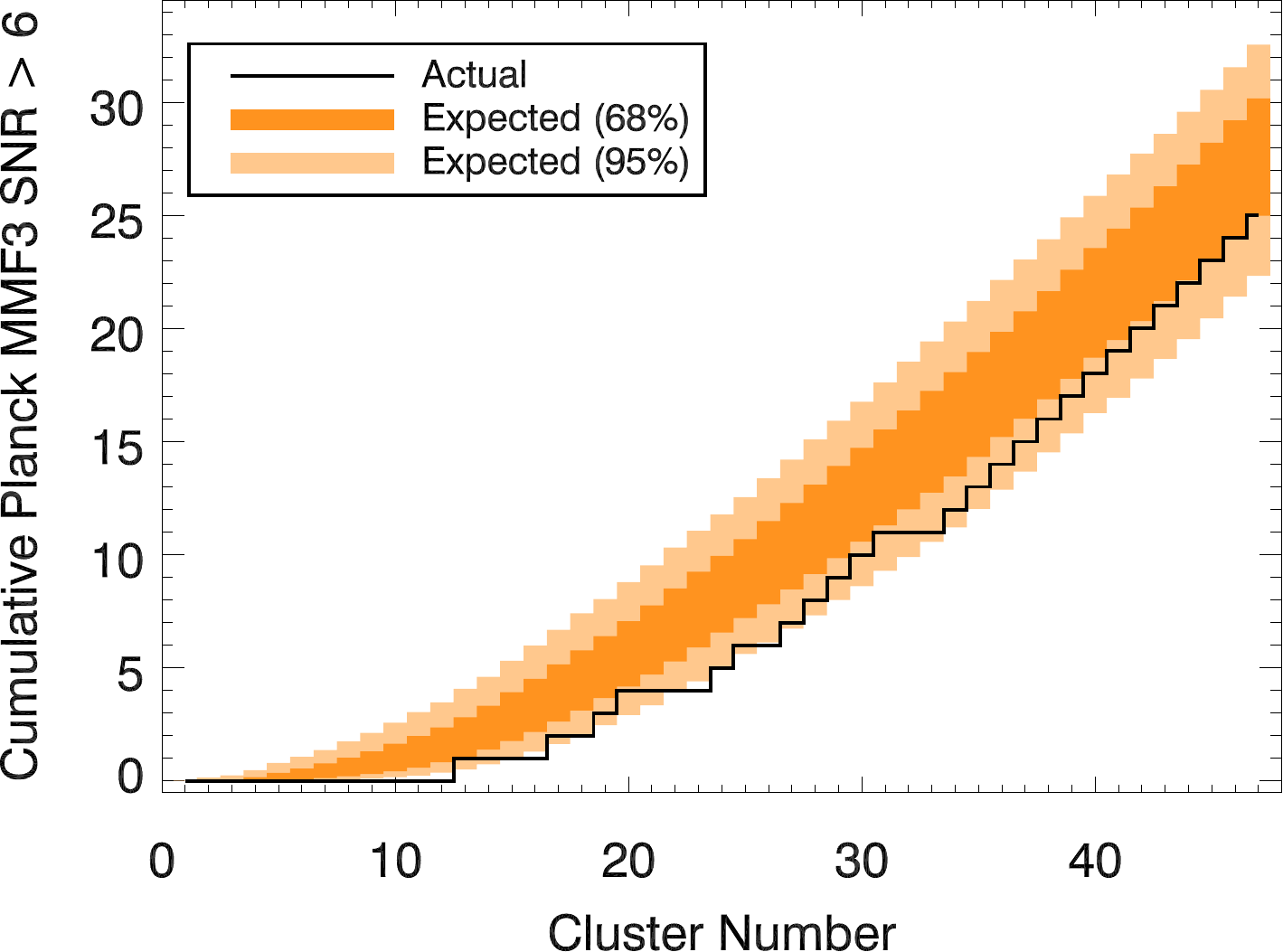}
      \caption{Left: The probability for each \boxs\ cluster to be detected with SNR $> 6$ by \planck\ using
        the MMF3 algorithm in ascending order of detection probability. The
        probability is computed using the Bolocam measurement of \ybig,
        rescaled according to the linear fit shown in Figure~\ref{fig:linmix_m500}. 
        Black diamonds denote the clusters actually
        detected by \planck\ and red triangles denote non-detections. The vertical bars
        represent the range of probabilities predicted from the Bolocam measurement of \ybig,
        with most of the uncertainty due to the intrinsic scatter in the linear
        model used to rescale the Bolocam measurements.
        Right: 68\% (dark orange) and 95\% (light orange)
        confidence regions for the total cumulative number of \planck\
        MMF3 clusters with SNR $> 6$  based on the detection probabilities given in
        the left plot. The actual cumulative number of \planck\ detections
        is given as a solid black line, and is consistent with, but
        somewhat low, compared to the
        predictions based on Bolocam.}
      \label{fig:planck_completeness}
      \vspace{12pt}
    \end{figure*}

    The \chandra\ and Bolocam values of \thetas\ and \yfive,
    rescaled to mimic the \xmm\ and \planck\ values as described
    in the previous paragraph,
    are then inserted into the \planck\
    SNR $=6$ completeness estimate to determine a detection probability for each 
    \boxs\ cluster
    (see Figure~\ref{fig:planck_completeness}).
    One subtlety is that the noise in the \planck\ maps is not uniform over the full sky, 
    and it is therefore necessary to account for this variation
    when calculating the detection probability for each \boxs\ cluster.
    Specifically, this variation is accounted for by comparing the
    noise RMS within the MILCA \ymap\ thumbnail centered on each
    cluster to the average noise RMS within the region of sky satisfying the
    cuts used for the \planck\ cluster analysis.
    In general, the local noise is within 5\% of the average,
    and the most extreme local noise deviation is 12\%.

    The left panel of Figure~\ref{fig:planck_completeness} shows the
    probability for every \boxs\ cluster to be detected by the \planck\ MMF3
    algorithm with a $\textrm{SNR} > 6$.
    There are no obvious outliers, with \planck\ detecting all of the clusters
    with a probability of $\sim 1$ and none of the clusters with a probability of $\sim 0$.
    To provide a quantitative test, a simulation was performed based on the 
    estimated detection probabilities. For each run of the simulation,
    a random value was drawn for each \boxs\ cluster
    based on the detection probability distribution for that cluster,
    and the total cumulative number of detections was computed.
    The simulation was repeated 10000 times, and the resulting
    68\% and 95\% confidence regions on the cumulative detections
    are plotted in the right panel of Figure~\ref{fig:planck_completeness}.
    The average number of detections in the simulations is 27.6, 
    and 16\% of the simulations result in fewer than the actual
    number of clusters detected by \planck, which is 25.

    This result provides a more extensive validation of the
    \planck\ completeness estimate,
    although \planck\ does detect slightly fewer clusters than
    expected given the Bolocam \yfive\ measurements.
    Such a shortfall could partially explain the tension seen between the CMB-derived
    and cluster-derived cosmological constraints 
    \citep{Planck2015_XXIV, Planck2015_XIII}.
    For example, \citet{Planck2015_XXIV} quantifies the level of tension in terms of a 
    cluster mass bias, with a value of $(1-b) = 0.58$ required to forge
    agreement.
    This is smaller than the true mass bias,
    with $(1-b) \simeq 0.7$--$0.8$
    found from lensing-based mass calibrations 
    \citep{Planck2015_XXIV, vonderLinden2014, Hoekstra2015},
    and the remaining 10--20\% difference is not well understood.
    Following this convention, the discrepancy between the predicted and actual
    number of \planck\ detections from the \boxs\ sample can be recast
    as an effective mass bias.
    In order for the average number of predicted detections to equal the actual 
    number of 25, the
    Bolocam \yfive\ measurements would need to be lower by a factor of $0.88 \pm 0.11$.
    Based on the \yfive/\mfive\ scaling relation derived in 
    \citet{Planck2013_XX}, 
    this corresponds to an effective mass bias factor
    of $(1-b) = 0.93 \pm 0.06$.
    This effective bias is multiplicative with the true mass bias,
    and would bring the \planck\
    cluster results into better agreement with the \planck\
    CMB results.
    
\section{Summary}
  \label{sec:summary}

  We fit A10 models to the \planck\ MILCA \ymap s using an \xmm-like prior
  on the value of \thetas, obtaining \ybig\ values
  consistent with those determined from the \planck\ MMF3
  algorithm using the same \thetas\ prior.
  We also derived \ybig\ from ground-based Bolocam
  observations, finding a Bolocam/\planck\ \ybig\ ratio of
  of $1.069 \pm 0.030$.
  This value is consistent with unity given calibration uncertainties
  and implies that Bolocam and \planck\ measure consistent SZ
  signals.
  Our results are in good agreement with previous comparisons
  between \planck\ and the ground-based AMI and CARMA-8 receivers,
  which yielded similar consistency

  We also performed joint fits to the Bolocam and \planck\ data, using a 
  gNFW model with the outer logarithmic slope $\beta$ allowed to
  vary with the other shape parameters fixed to the A10 values.
  These fits produce average values of $\beta = 6.13 \pm 0.16 \pm 0.76$
  and \yfive/\ybig\ $= 0.66 \pm 0.02 \pm 0.10$, which are
  in good agreement with recent simulations for clusters 
  matching the masses and redshifts of the \boxs\ sample.
  Compared to simulations,
  our data are also consistent with the trend of increasing $\beta$ with
  increasing cluster mass, but they do not
  reproduce the relatively strong trend of increasing $\beta$ with
  increasing redshift, likely due to selection effects in the \boxs\ sample.
  Previous SZ measurements of $\beta$ and \yfive/\ybig\
  indicate lower values
  than our results, although some or all of this difference may be 
  due to a combination of different median masses and redshifts
  within those samples, different
  mass measurements used to set the cluster radial scale,
  and/or measurement noise.

  Using Bolocam measurements of \yfive\ and \chandra\ measurements
  of \thetas, both rescaled to account for systematic differences
  relative to \planck\ measurements of \yfive\ and \xmm\
  measurements of \thetas, 
  we compute the detection probability for each \boxs\ cluster
  using the publicly available \planck\ completeness estimate.
  We estimate that \planck\ should
  detect an average of 27.6 \boxs\ clusters above the MMF3 SNR limit
  for the cosmology sample, a value that is within $\simeq 1 \sigma$
  of the actual number of \planck\ detections, which is 25.
  Our results therefore provide a further validation of the \planck\
  completeness estimate.
  Taking the small discrepancy at face value, however,
  may suggest that \planck\ detects fewer clusters than expected.
  Translated to an effective mass bias, this discrepancy
  yields $(1 - b) = 0.93 \pm 0.06$.
  This effective mass bias is multiplicative with the true mass bias
  of $(1-b) \simeq 0.7$--0.8 determined from lensing measurements, and
  would partially account for the 
  difference between the \planck\ cluster-derived
  and CMB-derived cosmological parameters
  that has not been explained by the lensing measurements
  \citep{Planck2015_XIII, Planck2015_XXIV}.

\section{Acknowledgments}

  We acknowledge the assistance of: Kathy Deniston, who provided effective
  administrative support at Caltech;
  James Bartlett and Jean-Baptiste Melin, who provided useful discussions;
  JS was supported by a NASA/ADAP award;
  MN, GP, and BW were supported by the Caltech Summer Research Connection
  program.
  SRS was supported by a NASA Earth and Space Science Fellowship
  and a generous donation from the Gordon and Betty Moore Foundation.

{\it Facilities:} \facility{Caltech Submillimeter Observatory}, \facility{{\it Planck}}, \facility{{\it Chandra}}.  

\bibliography{ms}
\bibliographystyle{aasjournal}

\end{document}